\documentclass[12pt,amsmath,amssymb,]{revtex4}
\usepackage{graphics}
\usepackage{amsmath}
\usepackage{epsfig}
\usepackage{epsf}
\usepackage{graphicx}
\usepackage{dcolumn}
\usepackage{bm}
\newcommand {\sla}[1]{ #1 \!\!\!/}

\begin{document}
\title{On the two-boson exchange corrections to  \\parity-violating elastic electron-proton scattering}

\date{\today}

\author{Hai Qing Zhou$^1$, Chung Wen Kao$^2$, Shin Nan Yang$^3$, and Keitaro Nagata$^{2,4}$ \\
$^1$Department of Physics,
Southeast University, NanJing 211189, China\\
$^2$Department of Physics, Chung-Yuan Christian University,\\
Chung-Li 32023, Taiwan\\
$^3$Department of Physics and Center for Theoretical Sciences,\\
National Taiwan University, Taipei 10617, Taiwan\\
$^4$Research Institute for Information Science and Education,\\
Hiroshima University, Higashi-Hiroshima 739-8521, Japan \\ }

\begin{abstract}
The details of the calculation of the two-boson exchange effects in
the parity-violating elastic $ep$ scattering within a simple
hadronic model, including both the nucleon and
$\Delta(1232)$-resonance intermediate states, are presented. We
examine the sensitivity of our results with respect to choice of
form factors. We emphasize the importance to use correct relations
relating $N\rightarrow \Delta$ and
$\Delta\rightarrow N$ transition vertex functions.   The $N\Delta$
Coulomb quadrupole transition is found to play important role at
 higher $Q^2\geq 3.0$ GeV$^2$. We also elucidate the relation between our
results and the well-known result on the $\gamma ZE$ effect given by
Marciano and Sirlin (MS). The effect of the nucleon contribution
$\delta_N$ to parity-asymmetry $A_{PV}$,  is found to be in general,
larger than the corresponding $\Delta$ contribution $\delta_\Delta$
except at extreme forward angles. The corrections to the extracted
values of the strange form factors $G^{s}_{E}+\beta G^{s}_{M}$ from
the HAPPEX, A4, and G0 data are also presented. The total TBE
corrections to the extracted values of $G^{s}_{E}+\beta G^{s}_{M}$
in recent experiments of HAPPEX  G0, and A4 are, depending on
kinematics, found to  be small except in a few cases where they
range from $-20.6\%$ to $48.3 \%$.
\end{abstract}

\maketitle

\section{Introduction}

One of the most intriguing questions in hadron structure  is the
possible existence of strangeness content in the proton  since
practically all constituent quark models employ only $u$ and $d$
quarks for light baryons. It was prompted by the EMC experiments
\cite{EMC89} which indicate that the amount of spin carried by the
strange quark pairs $s\bar s$ is comparable to that carried by the
$u$ and $d$ quarks and polarized opposite to the nucleon spin.
Similar conclusion  was also drawn from elastic $\nu p$ scattering
\cite{Ahrens87} and  theoretical analysis of $\pi N$ sigma term
\cite{Donoghue86}. A few other experiments have since been proposed
\cite{Ellis01}, including the  excess of $\phi$ production in $p\bar
p$ annihilation \cite{Amsler98},  $\Lambda$ polarization in
deep-inelastic neutrino scattering \cite{Ellis96,Nomad00}, and
double polarizations in photo- and electroproduction of $\phi$ meson
\cite{Titov97}  scheduled at SPring8 for 2010 \cite{WCChang09}, and
the parity-violating electron-proton scattering.

Parity-violating $ep$ scatterings was first suggested as a unique
probe to extract proton strange form factors by Kaplan and Manohar
\cite{Kaplan88} from measuring the  parity-violating asymmetry
$A_{PV}=(\sigma_R-\sigma_L)/(\sigma_R+\sigma_L)$ with polarized
electrons, where $\sigma_{R(L)}$ is the cross section with a
right-handed (left-handed)  electron. The asymmetry arises from the
interference of weak and electromagnetic amplitudes. Weak neutral
current elastic scattering is mediated by the $Z$-boson exchange and
measures form factors which are sensitive to a different linear
combination of the three light quark distributions.  When combined
with proton and neutron electromagnetic form factors and with the
use of charge symmetry, the strange electric and magnetic form
factors, $G^s_E$ and $G^s_M$, can then be determined
\cite{Kaplan88}. Since this is a rather clean technique to access
the charge and magnetization distributions of the strange quark
within nucleons, four experimental programs SAMPLE \cite{SAMPLE},
HAPPEX \cite{HAPPEX}, A4 \cite{A4}, and G0 \cite{G0} have been
designed to measure this important quantity, which is small and
ranges from  1 to 100 ppm. These experiments have been able to reach
a precision of $\delta{A_{PV}} \sim 0.1$ ppm. Several global
analyses have been performed \cite{Young06,Liu07,Pate08} and found
that the electric and magnetic strange form factors are quite small
with considerable error bars. Accordingly, greater effort  to reduce
theoretical uncertainty is needed in order to arrive at a more
reliable interpretation of experiments.

Leading order radiative corrections to $A_{PV}$, including the box
diagrams  Fig. 1(d) and other diagrams, have been extensively
studied \cite{Wheater82, Marciano83,Marciano84,Musolf90} and widely
used in the global analyses in \cite{Young06, Liu07, Pate08}. Among
those corrections, the interference between $\gamma Z$ exchange
($\gamma ZE$) of Fig. 1(d) with  Fig. 1(a), was evaluated  within
the zero momentum transfer approximation, i.e., $Q^2=0$. The first
calculation beyond the $Q^2=0$ approximation was done in
\cite{afanasev05} where the contribution of the interference of the
two-photon exchange ($2\gamma E$) process of Fig. 1(c) with diagram
of Figs. 1(a) and 1(b) to $A_{PV}$, was evaluated in a partonic
approach using GPDs. It was prompted by the fact that such a parton
model calculation of the $2\gamma E$ effect \cite{Chen04} was
arguably able to quantitatively resolve the discrepancy between the
measurements of the proton electric to magnetic form factor ratio
$R=\mu_pG_E/G_M$, where $\mu_p=2.79$, from Rosenbluth technique and
polarization transfer technique at high momentum-transfer-squared
$Q^2$ \cite{Jones00}. It was found \cite{afanasev05} that the
$2\gamma E$ correction to $A_{PV}$ is both $Q^2$ and $\epsilon$
dependent, and can reach several percent in certain kinematics,
becoming comparable in size with existing experimental measurements
of strange-quark effects in the proton neutral weak current.
However, the partonic calculations of \cite{afanasev05,Chen04} are
reliable only for $Q^2$ large comparable to a typical hadronic
scale, while all current experiments \cite{SAMPLE,HAPPEX,A4,G0} have
been performed at lower $Q^2$ values.

The two-boson exchange (TBE) corrections to $A_{PV}$, namely, the
contributions of the interference of the two-photon exchange
($2\gamma E$) process of Fig. 1(c) with diagram of Figs. 1(a) and
1(b) to $A_{PV}$, and that between the $\gamma Z$ exchange  of Fig.
1(d) with Fig. 1(a), were investigated in a hadronic model first
with only intermediate states restricted to elastic nucleon states
in \cite{zhou07,tjon08}. This hadronic model was developed in
\cite{Blunden03} to evaluate the $2\gamma E$ contribution to the
ratio $R$. The advantage of such a hadronic approach
\cite{Blunden03} is that it is applicable to low $Q^2$ region and
the results obtained are in agreement with the partonic calculation
of \cite{Chen04}. It is found \cite{zhou07,tjon08} that both the the
$2\gamma E$ and $\gamma ZE$ corrections to $A_{PV}$ depends strongly
on $Q^2$ and $\epsilon$, and can reach a few percent and are
comparable in size with the current experimental measurements of
strange quark effects in the proton weak neutral current and their
combined effects on the extracted values of $G_E^s+\beta G_M^s$ can
be as large as $-40 \%$ in certain kinematics. It was further found
\cite{tjon08} that the results show some sensitivity on whether a
monopole or dipole form is assumed for the nucleon form factors.

Recently, the  hadronic calculations on the TBE effects
\cite{zhou07,tjon08} were extended to include $\Delta(1232)$
resonance in the intermediate states \cite{Nagata09,Tjon09} since
$\Delta(1232)$ is known to play  a dominant role in low-energy
hadron physics \cite{pascal07}. Both calculations show that the
interplay between the nucleon and $\Delta$ contributions depend
strongly on the kinematics. However, there are discrepancies in the
size of the total TBE corrections due to the use of different vertex
relation relating the vertices of $\gamma N\rightarrow\Delta$ and
$\gamma\Delta\rightarrow N$, the strength of  the Coulomb
quardrupole excitation of the $\Delta$, and the $\Delta$ form
factors.

\begin{figure}[t]
\centerline{\epsfxsize 1.5 truein\epsfbox{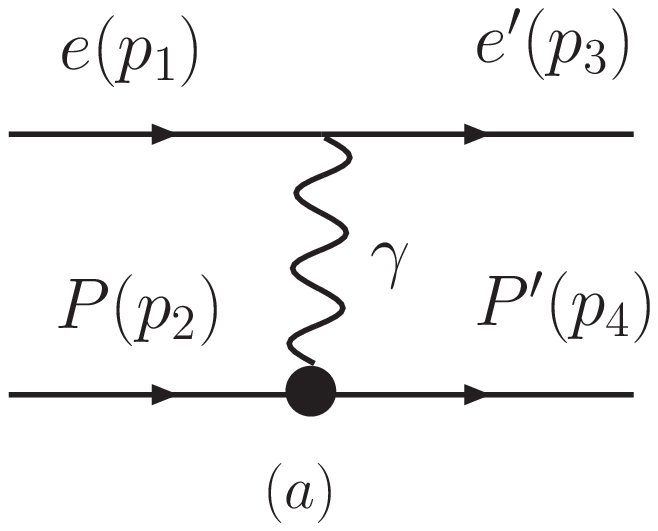} \epsfxsize
1.5 truein\epsfbox{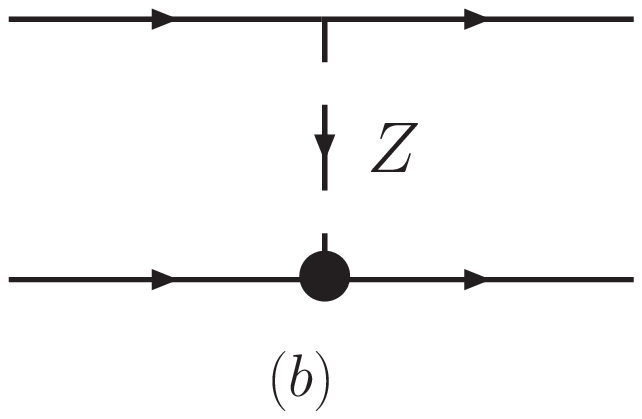}} \centerline{\epsfxsize
1.5truein\epsfbox{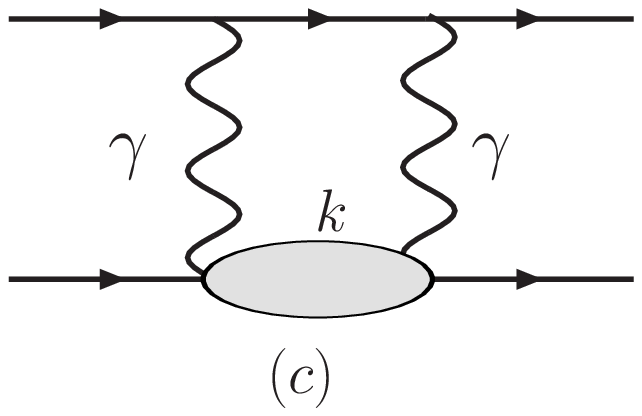} \epsfxsize 1.5
truein\epsfbox{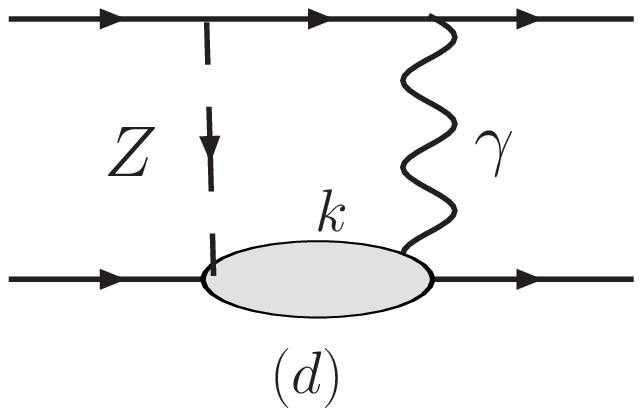}} \caption{(a) One-photon exchange, (b)
$Z$-boson exchange, (c) Two-photon exchange, and (d) $\gamma
Z$-exchange diagrams for elastic {\it ep} scattering. Corresponding
cross-box diagrams are implied.}
\end{figure}

In this paper, we give the details of our hadronic model
calculations \cite{zhou07,Nagata09} of the $2\gamma E$ and $\gamma
ZE$ corrections to $A_{PV}$ and present a more extensive results of
our calculation. In particular we analyze in details the difference
between our calculations and those of Ref. \cite{tjon08,Tjon09}. In
addition, we demonstrate explicitly that our results do recover the
results of \cite{Marciano84} in the limit of $Q^2=0$.

This article is organized as follows. The formalism for
parity-violating electron-proton is given in Section II. The details
of our calculation of the $\gamma ZE$ and $2\gamma E$ box diagrams
in a simple hadronic model are presented in Section III. The
numerical results of the above calculations and the impacts of our
results on the extraction of the strange form factors and the weak
charge of the proton are discussed in Section IV. In section V, we
summarize our work.
\section{Parity-violating electron-proton elastic scattering}

In this section we first briefly present the formulation of  the
parity-violating electron-proton elastic scattering within one-boson
exchange (OBE) approximation and the corresponding procedure to
extract the proton strange form factors.   We then  go beyond the
OBE framework and discuss radiative corrections.

\subsection{Parity-violating $ep$ scattering within one-boson-exchange approximation}

The  OBE diagrams of the elastic electron-proton scattering,
$e(p_1)+p(p_2)\rightarrow  e(p_3)+p(p_4)$, include one-photon
exchange $(1\gamma E)$ and  one-Z-boson exchange $(1Z E)$ diagrams,
as shown in   Fig. 1(a) and 1(b), respectively. At hadron level,
the couplings of the photon and $Z$-boson with the proton are given
as
\begin{eqnarray}
\langle p'|J^{Z}_\mu|p\rangle
&=&\overline{u}(p')\left[F^{Z,p}_{1}(q^2)\gamma_\mu+F^{Z,p}_{2}(q^2)\frac{i\sigma_{\mu\nu}}{2M_{N}}q^\nu+G^Z_A
(q^2)\gamma_\mu\gamma_5\right]u(p), \nonumber \\
\langle p'|J^{em}_\mu|p\rangle
&=&\overline{u}(p')\left[F^{\gamma,p}_{1}(q^2)\gamma_\mu+F^{\gamma,p}_{2}(q^2)
\frac{i\sigma_{\mu\nu}}{2M_{N}}q^\nu\right]u(p), \label{formfactors}
\end{eqnarray}
where $M_{N}$ is the proton mass and $q=p'-p$. $F^{\gamma/Z}_{1,2}$
and $G^Z_A$ are the proton electromagnetic/neutral weak current and
axial form factors, respectively. The Sachs form factors are defined
as
\begin{equation}
G_{E}^{\gamma/Z}=F_{1}^{\gamma/Z}-\tau F_{2}^{\gamma/Z},\,\,\,
G_{M}^{\gamma/Z}=F_{1}^{\gamma/Z}+F_{2}^{\gamma/Z},
\end{equation}
where $\tau=\frac{Q^2}{4M^2_{N}}$ with $Q^2=-q^2$. The OBE diagrams
of Figs. 1(a) and 1(b) are given in terms of the matrix elements of
the electromagnetic and neutral weak currents
\begin{eqnarray}
M^{(a)}&=&-i\overline{u}(p_3)(-ie\gamma^{\mu})
u(p_1)\frac{-i}{q^2+i\varepsilon}\overline{u}(p_4)
\Gamma^\gamma_{\mu}(p_{4},p_{2})u(p_2), \nonumber\\
M^{(b)}&=&-i\overline{u}(p_3)\left(\frac{ig\gamma^\mu}{4\cos\theta_{W}}\right)[(-1+4\sin^{2}\theta_{W})+\gamma_{5}]u(p_1)
\frac{-i}{q^2-M_Z^2+i\varepsilon} \nonumber\\
&& \overline{u}(p_4)\Gamma^Z_\mu(p_{4},p_{2}) u(p_2),\label{Born}
\end{eqnarray}
where $\Gamma^{\gamma}_\mu(p',p) = ie\langle
N(p')|J^{em}_\mu|N(p)\rangle$, $\Gamma^Z_\mu(p',p) =
(ig/4\cos\theta_{W})\langle N(p')|J^{Z}_\mu|N(p)\rangle$.
$g=e/\sin\theta_{W}$ is the weak coupling constant  with
$\theta_{W}$ the Weinberg weak mixing angle, and $M_Z$ the $Z$-boson
mass. The parity asymmetry in OBE approximation arises from  the
interference of $M^{(a)}$ and $M^{(b)}$.
Straightforward calculation, leads to the following expression of
parity-asymmetry in OBE approximation   in terms of the form factors
defined in Eq. (\ref{formfactors})
\begin{equation}
\begin{split}
A_{PV}&=-\frac{G_F
Q^2}{4\pi\alpha_{em}\sqrt{2}}\frac{A_E+A_M+A_A}{\bigl[\epsilon
(G^{\gamma,p}_{E})^2+ \tau (G^{\gamma,p}_{M})^2\bigr]},\\
A_E&=\epsilon G_{E}^{Z,p} G_{E}^{\gamma,p},
~~A_M=\tau G_{M}^{Z,p} G_{M}^{\gamma,p}, \\
A_A&=-(1-4\sin^{2}\theta_{W}) \sqrt{\tau (1+\tau) (1-\epsilon^2)}
G_A^Z G_{M}^{\gamma,p},
\end{split}
\label{AEMA}
\end{equation}
where $\epsilon \equiv [1+2(1+\tau)\tan^2\theta_{Lab}/2]^{-1}$ and
$\theta_{Lab}$  the scattering angle of the electron in the
laboratory frame. $G_{F}= \sqrt{2}g^2/8M_{W}^2=
\pi\alpha_{em}/(\sqrt{2}M_{Z}^{2}\cos^{2}\theta_{W}\sin^{2}\theta_{W})=
1.166\times 10^{-5} GeV^{-2}$ is the Fermi constant and
$\alpha_{em}= e^2/4\pi $   the fine structure constant.

To extract the strange form factors from Eq. (\ref{AEMA}), one needs
to make flavor decompositions of the form factors
$G^{\gamma,p}_{E,M}$ and $G^{Z,p}_{E,M}$. In the standard model, the
electromagnetic current and the neutral weak current are given as
\begin{equation}
J_{\mu}^{em}=\sum_{f=u,d,s} Q_{f}
\bar{q}_{f}\gamma_{\mu}q_{f},\,\,\,
J_{\mu}^{Z}=\sum_{f=u,d,s}\bar{q}_{f}\gamma_{\mu}(g^{f}_{V}+g^{f}_{A}\gamma_{5})q_{f},
\label{quarkcurrent}
\end{equation}
where   \cite{Musolf92},
\begin{eqnarray}
g^{e}_{V}&=&-1+4\sin^{2}\theta_{W},\,\,g^{e}_{A} = +1,\nonumber \\
g^{u}_{V}&=&1-\frac{8}{3}\sin^{2}\theta_{W},\,\,g^{u}_{A} = -1,\nonumber \\
g^{d}_{V}&=&-1+\frac{4}{3}\sin^{2}\theta_{W},\,\,g^{d}_{A} = +1,\nonumber \\
g^{s}_{V}&=&-1+\frac{4}{3}\sin^{2}\theta_{W},\,\,g^{s}_{A} = -1.
\label{gtree}
\end{eqnarray}
From Eqs. (\ref{formfactors},\ref{quarkcurrent},\ref{gtree}), one
obtains
\begin{eqnarray}
G^{\gamma,p}_{E,M}&=&\frac{2}{3}G^{u/p}_{E,M}-\frac{1}{3}G^{d/p}_{E,M}-\frac{1}{3}G^{s/p}_{E,M},
\nonumber \\
G^{\gamma,n}_{E,M}&=&\frac{2}{3}G^{u/n}_{E,M}-\frac{1}{3}G^{d/n}_{E,M}-\frac{1}{3}G^{s/n}_{E,M},
\nonumber \\
G^{Z,p}_{E,M}&=&(1-\frac{8}{3}\sin^{2}\theta_{W})G^{u/p}_{E,M}+(-1+\frac{4}{3}\sin^{2}\theta_{W})
G^{d/p}_{E,M}+(-1+\frac{4}{3}\sin^{2}\theta_{W})G^{s/p}_{E,M},
\label{NffinQff}
\end{eqnarray}
where $G^{q_{f}/p}_{E,M}$ are defined as follows
\begin{equation}
\langle p(p')|\bar{q}_{f}\gamma_{\mu}q_{f}|p(p)\rangle
=\overline{u}_{p}(p')\left[F^{q_{f}/p}_{1}(q^2)\gamma_\mu+F^{q_{f}/p}_{2}(q^2)\frac{i\sigma_{\mu\nu}}{2M}q^{\nu}\right]u_{p}(p),
\end{equation}
and
\begin{equation}
G_{E}^{q_{f}/p}=F_{1}^{q_{f}/p}-\tau F_{2}^{q_{f}/p},\,\,\,
G_{M}^{q_{f}/p}=F_{1}^{q_{f}/p}+F_{2}^{q_{f}/p}.
\end{equation}

If charge symmetry is assumed, i.e., the distribution of the $u$
quarks in the proton is the same as that of the $d$ quarks in the
neutron,  then one has $G^{u/p}_{E,M}=G^{d/n}_{E,M}$,
$G^{d/p}_{E,M}=G^{u/n}_{E.M}$, and
$G^{s/p}_{E,M}=G^{s/n}_{E,M}=G^{s}_{E,M}$ such that we can express
the neutron electromagnetic form factors as
\begin{equation}
G^{\gamma,n}_{E,M}=\frac{2}{3}G^{d/p}_{E,M}-\frac{1}{3}G^{u/p}_{E,M}-\frac{1}{3}G^{s}_{E,M}.
\label{nEMff}
\end{equation}
Combining the first equation of Eq. (\ref{NffinQff}) and Eq.
(\ref{nEMff}) leads to the following two relations
\begin{equation}
G^{u/p}_{E,M}=2G^{\gamma,p}_{E,M}+G^{\gamma,n}_{E,M}+G^{s}_{E,M},\,\,\,
G^{d/p}_{E,M}=G^{\gamma,p}_{E,M}+2G^{\gamma,n}_{E,M}+G^{s}_{E,M}.
\label{Gud}
\end{equation}
Putting the above two relations of Eq. (\ref{Gud}) back to the last
relation in Eq. (\ref{NffinQff}), the neutral weak form factors can
be expressed in terms of the electromagnetic form factors of the
proton and  neutron, and the strange form factors
\begin{equation}
G_{E,M}^{Z,p}=(1-4\sin^{2}\theta_{W})G^{\gamma,p}_{E,M}-G^{\gamma,n}_{E,M}-G^{s}_{E,M}.
\label{treeGZ}
\end{equation}
With Eq. (\ref{treeGZ}),   the parity asymmetry $A_{PV}$ of Eq.
(\ref{AEMA}) can be rewritten as,
\begin{eqnarray}
&&A_{PV}=A_1+A_2+A_3, \nonumber\\
&& A_{1}= -a\left[(1-4\sin^{2}\theta_{W})-\frac{\epsilon
G^{\gamma,p}_{E}G^{\gamma,n}_{E}
+\tau G^{\gamma,p}_{M}G^{\gamma,n}_{M}}{\epsilon(G^{\gamma,p}_ {E})^2+\tau(G^{\gamma,p}_{M})^2}\right],\nonumber \\
&& A_{2}= a\frac{\epsilon G^{\gamma,p}_{E}G^{s}_{E}
+\tau G^{\gamma,p}_{M}G^{s}_{M}}{\epsilon(G^{\gamma,p}_{E})^2+\tau(G^{\gamma,p}_{M})^2},\nonumber \\
&&A_{3}=a(1-4\sin^{2}\theta_{W})\frac{\epsilon'
G^{\gamma,p}_{M}G_{A}^{Z}}
{\epsilon(G^{\gamma,p}_{E})^2+\tau(G^{\gamma,p}_{M})^2},
\label{TreeA123}
\end{eqnarray}
where $a= G_{F}Q^2/4\pi\alpha_{em}\sqrt{2}$ and
$\epsilon'=\sqrt{\tau (1+\tau) (1-\epsilon^2)}$. The electromagnetic
form factors $G^{\gamma,p}_{E,M}$ and $G^{\gamma,n}_{E,M}$ can be
extracted from the elastic electron scattering from proton and
deuteron (for the neutron),  and the axial form factor $G^{Z}_{A}$
can be extracted from the pion photoproduction \cite{Bernard02}.
Accordingly, one can extract $A_{2}$ from $A_{PV}$ to obtain the
strange form factors $G^{s}_{E}+\beta G^{s}_{M}$ from $A_{2}$, with
$\beta= \tau G^{\gamma,p}_{M}/\epsilon G^{\gamma,p}_{E}$, if
radiative corrections can be neglected.

To take charge symmetry breaking effect into account, one may simply
replace Eq. (\ref{treeGZ}) with
\begin{equation}
G^{Z}_{E,M}=(1-4\sin^{2}\theta_{W})G^{\gamma,p}_{E,M}-G^{\gamma,n}_{E,M}-G^{s}_{E,M}-G^{CSB}_{E,M},
\end{equation}
where
$G^{CSB}_{E,M}=\frac{2}{3}\left(G^{d,p}_{E,M}-G^{u,n}_{E,M}\right)-\frac{1}{3}\left(G^{u,p}_{E,M}
-G^{d,n}_{E,M}\right)$ and the extraction formula of Eq.
(\ref{TreeA123}) remains unchanged except  $G^{s}_{E,M}$ be replaced
by $\tilde{G}^{s}_{E,M}=G^{s}_{E,M}+G^{CSB}_{E,M}$. $G^{CSB}_{E,M}$
have been estimated in the constituent quark model \cite{Pollock95,
Miller98}, light-cone meson-baryon model \cite{Ma97}, and chiral
perturbation theory ($\chi$PT) \cite{Lewis99} with low-energy
constants extracted from resonance saturation  \cite{Kubis06}.

\subsection{Radiative corrections to the parity-violating $ep$ scattering}
Since the value of $A_2$ in Eq. (\ref{TreeA123}) is just about a few
percent of $A_1$, it is  not possible to neglect the electroweak
radiative corrections, which is of order ${\cal O}(\alpha_{em})$, to
obtain accurate information of the strange form factors of the
proton. This is the reason why high precision measurements and
precise knowledge of the radiative corrections are required  to
obtain reliable extraction of the strange form factors from $ep$
scattering.

The complete ${\cal O}(\alpha_{em})$ radiative corrections to
$A_{PV}$ derive from several different sources such as vertex
corrections, self-energy insertions of the fermions and gauge
bosons, $\gamma Z$ mixing, wave function renormalization, two-boson
exchange, besides the inelastic bremsstrahlung. They have been
extensively studied \cite{Wheater82,
Marciano83,Marciano84,Musolf90}. The radiative corrections to
$A_{PV}$ have been  conventionally taken into account by expressing
$A_{PV}$ in following form \cite{Musolf92}
\begin{eqnarray}
&&A_{PV}(\rho,\kappa)=A_1+A_2+A_3, \nonumber\\
&& A_{1}= -a\rho\left[(1-4\kappa\sin^{2}\theta_{W})-\frac{\epsilon
G^{\gamma,p}_{E}G^{\gamma,n}_{E}
+\tau G^{\gamma,p}_{M}G^{\gamma,n}_{M}}{\epsilon(G^{\gamma,p}_ {E})^2+\tau(G^{\gamma,p}_{M})^2}\right],\nonumber \\
&& A_{2}= a\rho\frac{\epsilon G^{\gamma,p}_{E}\tilde{G}^{s}_{E}
+\tau G^{\gamma,p}_{M}\tilde{G}^{s}_{M}}{\epsilon(G^{\gamma,p}_{E})^2+\tau(G^{\gamma,p}_{M})^2},\nonumber \\
&&A_{3}=a(1-4\sin^{2}\theta_{W})\frac{\epsilon'
G^{\gamma,p}_{M}G_{A}^{Z}}
{\epsilon(G^{\gamma,p}_{E})^2+\tau(G^{\gamma,p}_{M})^2}.
\label{A123}
\end{eqnarray}
When the parameters $\rho$ and $\kappa$   equal one, Eq.
(\ref{A123}) reduces to  Eq. (\ref{AEMA}), and one recovers  the
tree approximation. The linear combination of the strange form
factors, $G^{s}_{E}+\beta G^{s}_{M}$, has been extracted from $A_2$
in Eq. (\ref{A123}). In this paper, we will restrict ourself to
corrections arising from TBE.

\section{The amplitudes of two-boson exchange diagrams}
In this section we evaluate the two-boson exchange  diagrams in a
simple hadronic model where the form factors are inserted as
regulators and only the   nucleon and $\Delta(1232)$ resonance
intermediate states are included. We present the details of the
calculation, including the explicit forms of the form factors and
the values of parameters employed. As in
\cite{Blunden03,zhou07,Nagata09}, we use package
FeynCalc~\cite{Feyncalc} and LoopTools~\cite{Looptools} to do the
analytical and numerical calculations, respectively.

\subsection{The amplitudes of $2\gamma E$ and $\gamma Z$ exchange box
diagrams}

Choosing the Feynman gauge and neglecting the electron mass $m_e$ in
the numerators, one can write down the amplitudes of box diagrams
Fig. 1(c) and Fig. 1(d) with the nucleon intermediate states as
\begin{eqnarray}
M^{(c,N)}&=&-i\int\frac{d^4k}{(2\pi)^4}\overline{u}(p_3)(-ie\gamma^{\mu})\frac{i(\sla{p}_1+\sla{p}_2-\sla{k})}
{(p_1+p_2-k)^2-m_e^2+i\varepsilon}(-ie\gamma^{\nu}) u(p_1)
\nonumber \\
&\times&\frac{-i}{(p_4-k)^2-\lambda^2+i\varepsilon}\frac{-i}{(k-p_2)^2-\lambda^2+i\varepsilon}
\overline{u}(p_4)
\Gamma^\gamma_{\mu}(p_{4},k)\frac{i(\sla{k}+M_{N})}
{k^2-M^2_{N}+i\varepsilon}\Gamma^\gamma_{\nu}(k,p_{2})u(p_2),\nonumber \\
M^{(d,N)}&=&-i\int\frac{d^4k}{(2\pi)^4}\overline{u}(p_3)(-ie\gamma^{\mu})\frac{i(\sla{p}_1+\sla{p}_2-\sla{k})}
{(p_1+p_2-k)^2-m_e^2+i\varepsilon}
(i\frac{g\gamma^\nu}{4\cos\theta_{W}}) \nonumber \\
&\times&[(-1+4\sin^{2}\theta_{W}]+\gamma_{5})u(p_1)
\frac{-i}{(p_4-k)^2-\lambda^2+i\varepsilon}
\frac{-i}{(k-p_2)^2-M_Z^2+i\varepsilon}\nonumber \\
&\times&\overline{u}(p_4)\Gamma^\gamma_{\mu}(p_{4},k)\frac{i(\sla{k}+M_{N})}
{k^2-M^2_{N}+i\varepsilon}\Gamma^Z_\nu(k,p_{2}) u(p_2).
\label{eq:diagram}
\end{eqnarray}
The amplitudes for the cross-box diagrams can be written down
similarly. Because the amplitudes in Eq. (\ref{eq:diagram}) are
infrared divergent, an infinitesimal photon mass $\lambda$ has been
introduced in the photon propagators to regulate the IR divergence.
As explained in \cite{zhou07}, in the soft photon limit, the box
diagrams and their corresponding bremsstrahlung cross section give
no correction to $A_{PV}$.   To go beyond the soft photon
approximation to estimate the corrections to $A_{PV}$, we calculate
the full amplitudes of $M^{(c,N)}$ and $M^{(d,N)}$ and subtract
$M^{(c,N)}_{soft}$ and $M^{(d,N)}_{soft}$ from their respective full
amplitude. The interferences between the remaining box diagrams and
the tree diagrams are then IR safe.

Similarly, amplitudes for the diagrams Fig. 1(c) and Fig. 1(d)  with
the $\Delta(1232)$ intermediate states can be written as follows
\begin{eqnarray}
M^{(c,\Delta)}&=&-i\int\frac{d^4k}{(2\pi)^4}\overline{u}(p_3)(-ie\gamma_{\mu})\frac{i(\sla{p}_1+\sla{p}_2-\sla{k})}
{(p_1+p_2-k)^2-m_e^2+i\varepsilon} \times(-ie\gamma_{\nu})
u(p_1)\frac{-i}{(p_4-k)^2+i\varepsilon}\nonumber \\
&\times&\frac{-i}{(k-p_2)^2+i\varepsilon} \overline{u}(p_4)
\Gamma^{\mu\alpha,\gamma}_{\Delta\rightarrow N}(k,p_{4}-k)
\frac{-i(\sla{k}+M_{\Delta})P_{\alpha\beta}^{3/2}(k)}
{k^2-M_{\Delta}^2+i\varepsilon}\Gamma^{\beta\nu,\gamma}_{N\rightarrow \Delta}(k,k-p_{2})u(p_2),\nonumber \\
M^{(d,\Delta)}&=&-i\int\frac{d^4k}{(2\pi)^4}\overline{u}(p_3)(-ie\gamma_{\mu})\frac{i(\sla{p}_1+\sla{p}_2-\sla{k})}
{(p_1+p_2-k)^2-m_e^2+i\varepsilon}
(i\frac{g\gamma_\nu}{4\cos\theta_{W}})\nonumber \\
&\times&((-1+4\sin^{2}\theta_{W})+\gamma_{5})u(p_1)
\frac{-i}{(p_4-k)^2+i\varepsilon}
\frac{-i}{(k-p_2)^2-M_Z^2+i\varepsilon}\nonumber \\
&\times&\overline{u}(p_4)\Gamma^{\mu\alpha,\gamma}_{\Delta\rightarrow
N}(k,p_{4}-k) \frac{-i(\sla{k}+M_{\Delta})P_{\alpha\beta}^{3/2}(k)}
{k^2-M_{\Delta}^2+i\varepsilon}\Gamma^{\beta\nu,Z}_{N\rightarrow
\Delta}(k,k-p_{2})u(p_2), \label{eq:DeltaDiagram}
\end{eqnarray}
where
\begin{equation}
P_{\alpha\beta}^{3/2}(k)=g_{\alpha\beta}-\frac{\gamma_{\alpha}\gamma_{\beta}}{3}
-\frac{(\sla{k}\gamma_{\alpha}k_{\beta}+k_{\alpha}\gamma_{\beta}\sla{k})}{3k^2},
\end{equation}
is the spin-3/2 projector.
The amplitudes in Eq. (\ref{eq:DeltaDiagram}) are IR finite because
when the four-momentum of the photon approaches zero the $\gamma
N\Delta$ vertices also approach zero. Therefore we do not need to
put $\lambda$ in Eq. (\ref{eq:DeltaDiagram}). The vertex functions
$\Gamma's$ for $\Delta\rightarrow N$ are defined by
\begin{eqnarray}
&&\overline{u}(p+q)\Gamma^{\mu\alpha,\gamma}_{\Delta\rightarrow
N}(p,q)u^\Delta_{\alpha}(p)=-ie\langle
N(p+q)|J^\mu_{em}|\Delta(p)\rangle,\nonumber \\
&&\overline{u}(p+q)\Gamma^{\mu\alpha,Z}_{\Delta\rightarrow
N}(p,q)u^\Delta_{\alpha}(p)=-ig\langle
N(p+q)|J^\mu_{Z}|\Delta(p)\rangle,
\end{eqnarray}
and similarly vertex functions for $N\rightarrow\Delta$ are defined
by
\begin{eqnarray}
&&\overline{u}^\Delta_{\beta}(p)\Gamma^{\beta\nu,\gamma}_{N\rightarrow
\Delta}(p,q)u(p-q)=-ie\langle \Delta(p)|J^\nu_{em}|N(p-q)\rangle,
\nonumber \\
&&\overline{u}^\Delta_{\beta}(p )\Gamma^{\beta\nu,Z}_{N\rightarrow
\Delta}(p,q)u(p-q)=-ig\langle \Delta(p)|J^\nu_{Z}|N(p-q)\rangle.
\end{eqnarray}
Note that $q's$ in $\Gamma^{\mu\alpha,\gamma/Z}_{\Delta\rightarrow
N}(p,q)$ and $\Gamma^{\beta\nu, \gamma/Z}_{N\rightarrow\Delta}(p,q)$
always correspond to  the {\it incoming} momentum of the photon ($Z$
boson), a convention used in  \cite{Blunden03}.

The relations between these vertex functions are
\begin{equation}
\Gamma^{\gamma}_{\Delta\rightarrow
N}(p,q)=-\gamma_{0}[\Gamma^{\gamma}_{N\rightarrow\Delta}(p,-q)]^{\dagger}\gamma_{0}
,\,\,\, \Gamma^{Z}_{\Delta\rightarrow
N}(p,q)=-\gamma_{0}[\Gamma^{Z}_{N\rightarrow\Delta}(p,-q)]^{\dagger}\gamma_{0}.
\label{ours}
\end{equation}
On the other hand, the following relations
\begin{equation}
\Gamma^{\gamma}_{\Delta\rightarrow
N}(p,q)=\gamma_{0}[\Gamma^{\gamma}_{N\rightarrow\Delta}(p,q)]^{\dagger}\gamma_{0}
,\,\,\, \Gamma^{Z}_{\Delta\rightarrow
N}(p,q)=\gamma_{0}[\Gamma^{Z}_{N\rightarrow\Delta}(p,q)]^{\dagger}\gamma_{0},
\label{theirs}
\end{equation}
are used in  \cite{Blunden03,Tjon09}. We consider Eq. (\ref{ours})
to be the correct one because it can be derived from the fact that
both of the electromagnetic and neutral weak currents are Hermitian.
The difference between Eq. (\ref{ours}) and
Eq. (\ref{theirs}) incurs  discrepancies between the results
obtained in \cite{Nagata09} and \cite{Tjon09} as will be discussed
later.

\subsection{Matrix elements of the electromagnetic and neutral weak currents\\
 between nucleon and $\Delta$}
Here we discuss the explicit forms of $\Gamma_{\Delta\rightarrow
N}^{\gamma}$ and $\Gamma_{\Delta\rightarrow N}^{Z}$. The matrix
elements of electromagnetic current between N and $\Delta$ is
written as \cite{kondra05}
\begin{eqnarray}
\langle N(p')|J^{em}_\mu|\Delta(p)\rangle
&=&\frac{1}{M_{N}^{2}}\overline{u}(p')
[g_{1}F^{(1)}_{\Delta}(q^2)(g^{\alpha}_{\mu}\sla{p}\sla{q}
-p_{\mu}\gamma^{\alpha}\sla{q}
-\gamma_{\mu}\gamma^{\alpha}p\cdot q+\gamma_{\mu}\sla{p}q^{\alpha})\nonumber \\
&+&g_{2}F^{(2)}_{\Delta}(q^2)(p_{\mu}q^{\alpha}
-p\cdot q g^{\alpha}_{\mu})\nonumber \\
&+&g_{3}F^{(3)}_{\Delta}(q^2)/M_{N}(q^2(p_{\mu}\gamma^{\alpha}-g^{\alpha}_{\mu}\sla{p})
+q_{\mu}(q^{\alpha}\sla{p}-\gamma^{\alpha}p\cdot
q))]\gamma_{5}T_{3}^{\dagger}u_{\alpha}(p), \label{GamaNDelta}
\end{eqnarray}
where $q=p'-p$ and $T_3$ is the third component of the $N\rightarrow
\Delta$ isospin transition  operator. $g_{i}$ are constants and
$F^{(i)}_{\Delta}(q^2=0)=1$. One has the following relation between
$G_{E,M,C}$, the transition form factors defined by Jones and
Scadron \cite{Jones73} and $g_{1},\,g_{2},\,g_{3}$:
\begin{eqnarray}
g_1&=&\frac{3}{2}\frac{M_{N}}{M_{\Delta}+M_{N}}(G_{M}(0)-G_{E}(0)),\nonumber\\
g_2&=&\frac{3}{2}\frac{M_{N}(M_{\Delta}+3M_{N})}{M_{\Delta}^2-M_{N}^2}G_{E}(0)+\frac{3}{2}\frac{M_{N}}
{M_{\Delta}+M_{N}}G_{M}(0),\nonumber \\
g_3&=&-\frac{3}{2}\frac{M_{N}^{2}}{M_{\Delta}(M_{\Delta}+M_{N})}\left(-\frac{M_{\Delta}+M_{N}}{M_{\Delta}-M_{N}}
G_{C}(0)+\frac{4M_{\Delta}^2}{(M_{\Delta}-M_{N})^{2}}G_{E}(0)\right)
\end{eqnarray}
We take $G_{M}(0)=3.02$ \cite{Tiator01}. $G_{E}(0)$ and $G_{C}(0)$
can be inferred from the relations $G_{E}(0)=-G_{M}(0)R_{EM}$ and
$G_{C}(0)=-[4M_{\Delta}^2/(M_{\Delta}^{2}-M_{N}^{2})]G_{M}(0)
R_{SM}$ with the experimentally determined values of $R_{EM}=-2.5\%$
\cite{Beck00,DMT} and $R_{SM}=-4.0\%$ \cite{MAID07}. We thus have
$G_{E}(0)=0.0755$ and $G_{C}(0) =1.1496$ and correspondingly,
$g_{1}=1.91$, $g_{2}=2.63$, and $g_3=1.57$. Note that the
normalization used in Eq. (\ref{GamaNDelta}) to define the couplings
constants $g_i's$ differs from that of \cite{kondra05, Tjon09} where
they used  $M_\Delta$ instead of $M_N$ everywhere in Eq.
(\ref{GamaNDelta}). With this normalization difference  taken into
account, the corresponding values of $g_i's$ used in  \cite{Tjon09}
would be $g_{1}=1.82$ and $g_{2}=2.81$, with
  $g_{3}$ varied from -0.44 to 1.28.  We note that, however, since all the
current experimental data for $R_{SM}$ extracted from experiments at
low $Q^2$'s as small as $Q^2=0.060$ GeV$^2$ \cite{Stave06} remain
negative, we will not consider the possibility of a negative value
of $g_3$. The difference between the values of $g_{3}$ used in our
calculation and \cite{Tjon09}   leads to considerable differences in
some of the results between these two calculations, if the vertex
relation of Eq. (\ref{ours}) is employed, as will be discusses in
the next section.

The neutral weak current can be decomposed into isovector and
isoscalar parts:
\begin{eqnarray}
J^{Z}_{\mu}&=&\alpha_V V^{3}_{\mu}+\beta_A A^{3}_{\mu}+\rm{isoscalar}\,\, \rm{terms},\nonumber \\
J^{em}_{\mu}&=&V^{3}_{\mu}+\rm{isoscalar}\,\, \rm{terms},
\end{eqnarray}
where the superscript "3" refers to the third component  in isospin
space,   $\alpha_V = (1-2\sin^2\theta_{W})/(2\cos\theta_{W})$ and
$\beta_A =-1/(2\cos\theta_{W})$. The isoscalar part does not
contribute to $N \rightarrow \Delta$ transition. The $Zp\Delta^{+}$
vertex contains both the vector and the axial-vector components. The
vector part takes the form
\begin{eqnarray}
\langle p(p')|J^{Z}_{\mu,V}|\Delta^{+}(p)\rangle
&=&\frac{1}{M_{N}^{2}}\overline{u}(p')
[\tilde{g}_{1}F^{(1)}_{\Delta}(q^2)(g^{\alpha}_{\mu}\sla{p}\sla{q}
-p_{\mu}\gamma^{\alpha}\sla{q}-\gamma_{\mu}\gamma^{\alpha}p\cdot q+\gamma_{\mu}\sla{p}q^{\alpha})\nonumber \\
&+&\tilde{g}_{2}F^{(2)}_{\Delta}(q^2)(p_{\mu}q^{\alpha}
-p\cdot q g^{\alpha}_{\mu})\nonumber \\
&+&\tilde{g}_{3}F^{(3)}_{\Delta}(q^2)/M_{N}(q^2(p_{\mu}\gamma^{\alpha}-g^{\alpha}_{\mu}\sla{p})
+q_{\mu}(q^{\alpha}\sla{p}-\gamma^{\alpha}p\cdot
q))]\gamma_{5}u_{\alpha}(p), \label{ZNDelta}
\end{eqnarray}
where $\tilde{g}_{i}'s$ and $g_{i}'s$ are related by
$\tilde{g}_{i}=\sqrt{2/3}\, \alpha_V g_{i}.$ Note that the factor
$\sqrt{\frac{2}{3}}$ comes  from isospin transition operator
$T_{3}^{\dagger}$. Thus we have $\tilde{g}_{1}=0.47$,
$\tilde{g}_{2}=0.658$ and $\tilde{g}_{3}=0.40$.

The axial-vector component of the  $Zp\Delta^{+}$ vertex is given as
\cite{Nagata09}
\begin{eqnarray}
\langle p(p')|J^{Z}_{\mu,A}|\Delta^{+}(p)\rangle &=& \frac{
1}{M_{N}^{2}}\overline{u}(p')
[h_{1}H^{(1)}_{\Delta}(q^2)(g^{\alpha}_{\mu}(p\cdot q)-p_{\mu}q^{\alpha})\nonumber \\
&+&h_{2}H^{(2)}_{\Delta}(q^2)/M_{N}^{2}(q^{\alpha}q_{\mu}\sla{p}\sla{q}
-(p\cdot q)\gamma^{\alpha}q_{\mu}\sla{q})
+h_{3}H^{(3)}_{\Delta}(q^2)((p\cdot q)\gamma^{\alpha}\gamma_{\mu}
-\sla{p}\gamma_{\mu}q^{\alpha}) \nonumber \\
&+&h_{4}H^{(4)}_{\Delta}(q^2)(g^{\alpha}_{\mu}p^{2}-p_{\mu}\gamma^{\alpha}\sla{p})]u^\Delta_{\alpha}(p),
\label{axial}
\end{eqnarray}
where $F^{(i)}_{\Delta}$ in Eq. (\ref{ZNDelta}) and
$H^{(i)}_{\Delta}$ in Eq. (\ref{axial}) are the vector and
axial-vector form factors, respectively. In the present
investigation, we will assume  that, for simplicity,  some of them
separately takes a common form for different couplings, i.e.,
$F^{(i)}_{\Delta}=F_{\Delta}(Q^2)$  and
$H^{(i)}_{\Delta}=H_{\Delta}(Q^2)$. In addition, both $F_{\Delta}$
and $H_{\Delta}$ are normalized to one at $Q^2=0$.

Only  the coupling constants  $h_{i}'s$ remain to be determined.
They can be obtained from the data of $\nu N\rightarrow \mu \Delta$.
Many experimental papers on neutrino induced $\Delta$ production
adopt the notation of Llewellyn-Smith \cite{smith} where the
$N\Delta$ transition induced by the weak charged axial-current is
written as
\begin{eqnarray}
\langle \Delta^{++}(p')|J^{W,A}_\mu|p(p)\rangle
&=&\overline{u}_{\alpha}(p')
[\frac{C^{A}_{3}(Q^2)}{M_{N}}(\sla{q}g^{\alpha}_{\mu}-q^{\alpha}\gamma_{\mu})
+\frac{C^{A}_{4}(Q^2)}{M^{2}_{N}}((p'\cdot q)g^{\alpha}_{\mu}-q^{\alpha}p'_{\mu})\nonumber \\
&+&C^{A}_{5}(Q^2)g^{\alpha}_{\mu}
+\frac{C^{A}_{6}(Q^2)}{M_{N}^{2}}p^{\alpha}q_{\mu}]u(p).
\label{smith}
\end{eqnarray}

The form factors in Eq. (\ref{axial}) can be related to the form
factors defined in Eq. (\ref{smith}) by performing a rotation in
isospace and assuming the nucleon and $\Delta$ are both on-shell.
The resulting relations are
\begin{eqnarray}
&h_{1}=&-\frac{\beta
C^{A}_{4}(0)}{\sqrt{3}}-\frac{2M_{N}}{M_{\Delta}}\,\frac{\beta
C^{A}_{3}(0)}{\sqrt{3}},\nonumber\\
&h_{3}=&\frac{M_{N}}{M_{\Delta}}\cdot\frac{\beta
C^{A}_{3}(0)}{\sqrt{3}},\,
h_{2}=-\frac{M_{N}^{2}}{M_{\Delta}(M_{\Delta}-M_{N})}\,\frac{\beta C^{A}_{6}(0)}{\sqrt{3}},\nonumber \\
&h_{4}=&\frac{M_{N}^{2}}{M_{\Delta}^{2}}\frac{\beta
C^{A}_{5}(0)}{\sqrt{3}}+\frac{M_{N}(M_{\Delta}-M_{N})}{M_{\Delta}^{2}}\,\frac{\beta
C^{A}_{3}(0)}{\sqrt{3}}.
\end{eqnarray}
According to \cite{Adler68} and \cite{Schreiner73}, $C_{3}^{A} =0$
and hence $h_{3} =0$. If we follow the weak pion production data of
\cite{Kitagaki90} and
  extrapolate the experimental result to $Q^2 =0$ \cite{Hemmert95}, then we  find
$C^{A}_{4}(0)=-0.8,\,\,\,C^{A}_{5}(0)=2.4,$ to obtain
$h_{1}=-0.263,\,\,h_{4}=-0.458$. The parameter $h_2$ cannot be
determined from the weak pion production. According to partial
conservation of axial current (PCAC), one has the following relation
\begin{equation}
C_{6}^{A}(Q^2)\approx \frac{M^2_{N}}{m_{\pi}^{2}+Q^2}C^{A}_{5}(Q^2),
\end{equation}
where $m_{\pi}$ is the pion mass. Hence one obtains
$C^{A}_{6}(0)\approx \frac{M^2_{N}}{m_{\pi}^{2}}  C^{A}_{5}(0)
\approx 107.7$ and the corresponding value for $h_2$ would be about
$-360.91$. Even with such a large value, we find its effect is tiny
($\le 10^{-16}$) and therefore we simply set $h_2=0$.

Note that the vertices $\gamma N\Delta$ and $ ZN\Delta$ in Eqs.
(\ref{GamaNDelta}, \ref{ZNDelta}, \ref{axial}) all satisfy the
constraints:
\begin{equation}
p_{\alpha}\Gamma^{\mu\alpha}_{\Delta\rightarrow
N}(p,q')=p_{\beta}\Gamma^{\beta\nu}_{N\rightarrow \Delta}(p,q')=0,
\end{equation}
for any $q'$, to eliminate the coupling of the unphysical spin-1/2
component of Ratria-Schwinger spinor \cite{Pascal99}. The
expressions in Eqs. ({\ref{ZNDelta}, \ref{axial}) have been written
in many different ways \cite{smith,Nath,Hemmert95} but only those
given here satisfy the above constraints.

In \cite{Tjon09}, different forms of the axial form factors are
employed. They obtain the matrix elements of $J^{Z}_{\mu,A}$ by
simply removing  $\gamma_5$ from Eq. (\ref{ZNDelta}) and write

\begin{eqnarray}
\Gamma^{\mu\alpha,Z}_{\Delta\rightarrow N}&=&
\frac{i}{2M_{\Delta}^2}[g_{1}^{A}(Q^2)[g^{\mu\alpha}\sla{p}\sla{q}-p^{\mu}\gamma^{\alpha}\sla{q}
-\gamma^{\mu}\gamma^{\alpha}(p\cdot q)+\gamma^{\mu}q^{\alpha}\sla{p}] \nonumber \\
&+&g_{2}^{A}(Q^2)[p^{\mu}q^{\alpha}-g^{\mu\alpha}(p\cdot q)]\nonumber \\
&+&\frac{g_{3}^{A}(Q^2)}{M_{\Delta}}
[q^2(p^{\mu}\gamma^{\alpha}-g^{\mu\alpha}\sla{p})-q^{\mu}(q^{\alpha}\sla{p}-\gamma^{\alpha}(p\cdot
q))]]. \label{ZNDeltaTjon}
\end{eqnarray}
It leads to only three form factors instead of four in Eq.
(\ref{axial}). The form factor $g_{3}^{A}(Q^2)$ of Eq.
(\ref{ZNDeltaTjon}) is required to have a pole at $Q^2=0$. However,
form factors defined in Eq. (\ref{axial}) are not required to have
such poles, and in our opinion, a more appropriate choice.

\subsection{Nucleon and $N\rightarrow\Delta$ form factors}
So far we have not specified the explicit forms of the nucleon and $
N\rightarrow \Delta$ form factors. In this article we adopt the
following two sets of the  form factors. The   set A is parametrized
as follows:
\begin{eqnarray}
G^{\gamma,p}_{E}&=&G^{\gamma,p}_{M}/\mu_p = G^{Z,p}_{E}/x =
G^{Z,p}_{M}/y
=\frac{\Lambda^4_{1}}{(Q^2+\Lambda^2_{1})^2},\nonumber \\
G^{Z,p}_A/z &=& \frac{\Lambda^4_{2}}{(Q^2+\Lambda^2_{2})^2},
\label{NFFA}
\end{eqnarray}
where  $x=G^{Z,p}_{E}(Q^2=0)$, $y=G^{Z,p}_{M}(Q^2=0)$ and
$z=G^{Z}_{A}(Q^2=0)$. We take   $\Lambda_{1} = 0.84\,$ GeV and
$\Lambda_{2}= 1.0\,$ GeV from the usual dipole form $G^{\gamma,p}_E
= 1/(1+Q^2/0.71)^2$, and $G^Z_A = G^Z_A(0)/(1+Q^2)^2$
\cite{HAPPEX,Beise}, with $Q$ given in unit of $GeV$, i.e., $c=1$, a
convention to be used hereafter.
 We determine $x, y, z$
from relations \cite{Beise}, $G^{Z,p}_{E,M} = \rho(1-4\kappa
\sin^2\theta_W)G^{\gamma,p}_{E,M}-\rho G^s_{E,M} -\rho
G^{\gamma,n}_{E,M}$ and $G^Z_A = -(1+R_A^{T=1})G_A+\sqrt{3}R_A^{T=0}
G_A^8+\Delta s$ at $Q^2=0$ point. The quantities $G_A^8$ and $\Delta
s$ refer to the $SU(3)$ isoscalar octet form factor and the strange
quark contribution to the nucleon spin, respectively. The $\rho,
\kappa$ and $R_{A}^{T=1}$ and $R_{A}^{T=0}$ are due to radiative
corrections. They lead to $x=0.076\pm0.00264$, $y=2.08 \pm
0.00813-G_M^s(0), z = -0.95^{+0.37}_{-0.36}+\Delta s(0)$. We fix $x
= 0.076$ and vary the values of $y, z, \Lambda_1,$ and $\Lambda_2$
to check the sensitivity of the results on the parameters and find
 little changes.

The forms of the $\gamma N\Delta$ and $ZN\Delta$ are taken to be
\begin{equation}
F_{\Delta}(Q^2)=\frac{\Lambda_1^4} {(Q^2+\Lambda^{2}_1)^2},\,\,\,
H_{\Delta}(Q^2)=\frac{\Lambda^{4}_2} {(Q^2+ {\Lambda}^{2}_2)^2}.
\label{NDFFA}
\end{equation}
 Variations of these cutoffs are found not to
affect the results significantly as well.

The form factors set A given in Eqs. (\ref{NFFA}) and (\ref{NDFFA})
do not describe well the existing data at large $Q^2$. For example,
the ratio of the proton electric to magnetic form factors $R=\mu_p
G_E/G_M$ has been found to deviate from one at large $Q^2$
\cite{Jones00}, while the form factors of Eq. (\ref{NFFA}) gives
$R=1$. Similarly, the $N\rightarrow\Delta$ transition form factors
have been measured and found to drop faster than $Q^{-4}$ at high
$Q^2$. More specifically, perturbative QCD predicts that at high
$Q^2$, the Jones-Scadron form factors scale as follows
\cite{Carlson86},
\begin{equation}
G_M(Q^2)\sim Q^{-4}, \,\,G_E(Q^2)\sim Q^{-4}, \,\,G_C(Q^2)\sim
Q^{-6},
\end{equation}
such that both $R_{EM}$
and $R_{SM}$ should approach some constants as $Q^2\rightarrow
\infty$. The $N\rightarrow\Delta$ transition form factors given in
Eq. (\ref{NDFFA}) clearly do not have the correct asymptotic
behavior at high $Q^2$. We try to take these into account by adding
  extra   factors to both  $G^{\gamma,p}_{E}$ and
$F^{(i)}_\Delta$ given in Eqs. (\ref{NFFA}) and (\ref{NDFFA}).
This leads to the following  more realistic form factors set B, with
$F^{(1,2)}_{\Delta}(Q^2)=F_{\Delta}(Q^2)$,

\begin{eqnarray}
G^{\gamma,p}_{M}/\mu_{p}&=&\left(\frac{\Lambda_{1}^{2}}{Q^2+\Lambda_{1}^{2}}\right)^{2},\,
G^{\gamma,p}_{E}=\left(\frac{\Lambda_{1}^{2}}{Q^2+\Lambda_{1}^{2}}\right)^{2}
\frac{\Lambda_{3}^2}{Q^2+\Lambda_{3}^{2}},\,
G^{Z,p}_{A}=\frac{\Lambda_{1}^{2}}{Q^2+\Lambda_{1}^{2}},\nonumber \\
F_{\Delta}(Q^2)&=&\left(\frac{\Lambda_{1}^2}{Q^2+\Lambda_{1}^{2}}\right)^{2}
\frac{\Lambda_{4}^2}{Q^2+\Lambda_{4}^{2}},\,\,\,
F^{(3)}_{\Delta}=F_{\Delta}(Q^2)\left(\frac{\Lambda_{5}^2}{Q^2+\Lambda_{5}^{2}}\right),\,\nonumber \\
H_{\Delta}&=&\left(\frac{\Lambda_{1}^{2}}{Q^2+\Lambda_{1}^{2}}\right)^{2}.
\label{FFB}
\end{eqnarray}
Fitting the data of  \cite{Jones00,Sato07} gives   $\Lambda_{3}=2.0$ GeV,    $\Lambda_{4} = \sqrt{2}$ GeV and
$\Lambda_{5}=0.5$ GeV. Note that when one evaluates the effect from
the box diagrams with $\Delta$ intermediate states, one still needs
to specify the choice of the nucleon form factors because  one still
receives contribution from the interference between $1\gamma E$ and
TBE box diagrams. Therefore each form factors set includes both of
nucleon and $ N\rightarrow\Delta$ form factors. We will discuss the
sensitivity of the results with respect to the use of these two
different sets of the form factors in Sec IV.

\section{Results and Discussions}
In this section, we present the results of the corrections of
$2\gamma E$ and $\gamma ZE$ to $A_{PV}$ in the simple hadronic model
described in the previous section. The  sensitivity with respect to
 different choices of parameters and form factors will be analyzed in details. The
influence of the TBE effects on the extracted values of the strange
form factors $G^{s}_{E}+\beta G^{s}_{M}$ is   discussed at the end.

\subsection{TBE Effects on $A_{PV}$}

As in \cite{zhou07,Nagata09}, we characterize the $2\gamma E$ and
$\gamma ZE$ corrections to $A_{PV}$  by $\delta$ defined as
\begin{equation}
A_{PV}(1\gamma+Z+2\gamma+\gamma
Z)=A_{PV}(1\gamma+Z)(1+\delta_{N}+\delta_{\Delta}),
\end{equation}
where $A_{PV}(1\gamma+Z) $ denotes the parity-violating asymmetry
arising from the interference between $1\gamma$ and $Z$-boson
exchange, i.e., Figs. 1(a) and 1(b) while
$A_{PV}(1\gamma+Z+2\gamma+\gamma Z)$ includes the effects of
$2\gamma E$ and $\gamma ZE$ with the nucleon and $\Delta(1232)$
intermediate states. $\delta_{N(\Delta)}$ represents the
contribution from the diagrams with the nucleon ($\Delta$-resonance)
 intermediate states, respectively.

\subsubsection{The TBE corrections from the nucleon intermediate states}

\begin{figure}[h,b,t]
\centerline{\epsfxsize 4.0 truein\epsfbox{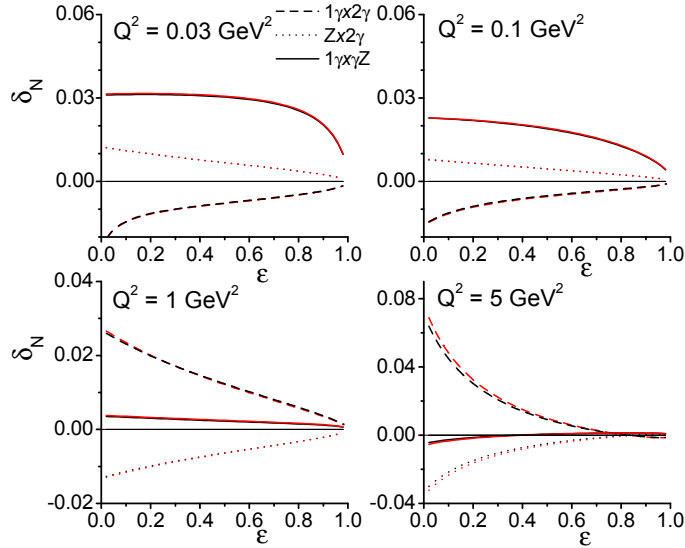}}
\caption{$2\gamma E$ and $\gamma ZE$ corrections to $A_{PV}$ with
nucleon intermediate states, as functions of $\epsilon$ from 0.1 to
0.99 at $Q^2=0.03,\, 0.1,\, 1.0,$ and $5.0\,$ GeV$^2$, respectively.
Dotted and dashed lines denote corrections coming only from
interferences of $1\gamma\times2\gamma$, and
 $Z\times2\gamma$ . The solid lines represent contributions of
 $\gamma\times\gamma Z$. The lines in red and black correspond to the results
 using the   form factors set A and  B,
respectively.} \label{N}
\end{figure}

We first present the results of $\delta_{N}$ as function of
$\epsilon$, the contributions of TBE diagrams with the nucleon
intermediate states, in Fig. {\ref{N}}. The effects of the
interferences between $1\gamma E$ and $2\gamma E$
($1\gamma\times2\gamma$), and those between $1ZE$ and $2\gamma E$
($Z\times2\gamma$) are represented by the dotted and dashed lines,
respectively, at  four different $Q^2$ values,
$Q^2$=0.03,\,\,0.1,\,\,1.0,\,\,5.0\,GeV$^{2}$. The interferences
between $1\gamma E$ and $\gamma ZE$ are given by the solid lines.
The lines in red correspond to the results obtained with form
factors set A while the black lines are associated with form factors
set B, as specified in Eqs. (\ref{NFFA}-\ref{FFB}) in the previous
section. We see little difference between   red and black curves in
Fig. {\ref{N}}, as  both form factors sets  A and B are of dipole or
higher order forms. On the contrary, in Fig. {\ref{Mono}} one finds
that at $Q^2=5.0$ GeV$^2$ the results using the monopole  form
factors, with cut-offs adjusted accordingly, are much smaller than
those obtained with sets A and B, as pointed out in \cite{tjon08}.

In Fig. {\ref{N}}, we see that both   $2\gamma E$ and $\gamma ZE$
effects strongly depend on $Q^2$ and $\epsilon$. The magnitude of
each contribution has its maximum at $\epsilon = 0$ and decrease to
zero when $\epsilon$ increases. One also sees that $1\gamma\times
2\gamma$ contribution always cancels the $Z\times2\gamma$
contribution and hence their sums are always small compared with the
size of each contribution, a feature also present in the partonic
calculation of Ref. \cite{afanasev05}. Another interesting fact is
that the magnitude of $\delta_N(1\gamma\times 2\gamma)$ is always
larger than $\delta_N(Z\times 2\gamma)$. For the $1\gamma\times
\gamma Z$ contribution, it decreases as $Q^2$ increases
 and dominates over $\delta_N(2\gamma E)$ at the backward directions when
$Q^2\le 0.1 \,$ GeV$^2$, but   reduces to about the same size as the
total $2\gamma E$ contribution at higher $Q^2$.
\begin{figure}[h,b,t]
\centerline{\epsfxsize 4.0 truein\epsfbox{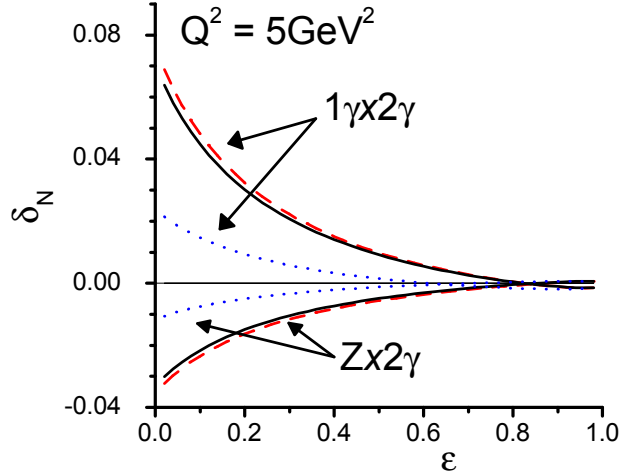}}
\caption{$1\gamma\times 2\gamma  $ and $ Z\times 2\gamma $
corrections to $A_{PV}$ with   nucleon intermediate states, as
functions of $\epsilon$ from 0.1 to 0.99 at $Q^2 = 5.0\,$ GeV$^2$.
Dotted lines denote the results obtained with  monopole type form
factors. Dashed and solid lines correspond to  results obtained with
form factors sets A and B, respectively.} \label{Mono}
\end{figure}

\subsubsection{The TBE corrections from the $\Delta$(1232) intermediate states}

\begin{figure}[h,b,t]
\centerline{\epsfxsize 4.0 truein\epsfbox{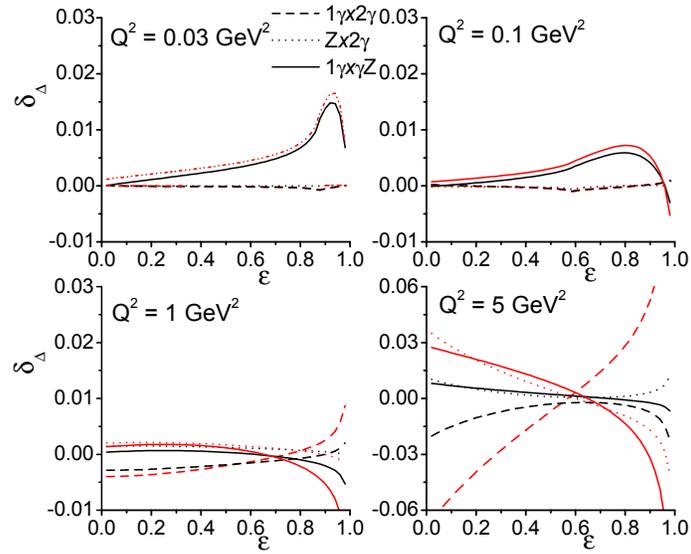}}
\caption{$2\gamma E$ and $\gamma ZE$ corrections to $A_{PV}$, with
$\Delta$(1232) intermediate states as functions of $\epsilon$   at
$Q^2=0.03,\, 0.1,\, 1.0,$ and $5.0\,$ GeV$^2$. Dotted and dashed
lines denote corrections coming only from the interference
$1\gamma\times2\gamma$, and $Z\times2\gamma$, respectively. The
solid lines represent the contribution of  $\gamma\times\gamma Z$.
The lines in red and black correspond to  results obtained with form
factors set A and  B, respectively. } \label{D}
\end{figure}

We continue to present our result of $\delta_{\Delta}$ which arises
from the TBE diagrams with the $\Delta(1232)$ intermediate states.
In Fig. {\ref{D}}, we show the $2\gamma E$ and $\gamma ZE$
corrections to $A_{PV}$ by plotting $\delta_{\Delta}\,\,vs. \,\,
\epsilon$ for both form factors sets
 A and B. Again, the red  and the black lines correspond to results obtained with
 form factors set A and  B, respectively. One immediately notices that they are very close
 to each other when $Q^2\le 0.1 $ GeV$^2$. However,
  difference begins to develop  when $Q^2$ reaches 1.0 GeV$^2$ at forward angles ($\epsilon\ge$ 0.8).
As $Q^2$  increases further, the difference between red and black
lines becomes more   pronounced even at small $\epsilon$ and at
$Q^2=5.0$ GeV$^2$, the discrepancy reaches more that $100\%$ in some
cases. The fact that $\delta_{\Delta}$ is more sensitive than
$\delta_{N}$ to the details of the form factors indicates that the
TBE diagrams with the $\Delta$ intermediate states are more strongly
dependent on the higher loop momentum than the diagrams with the
nucleon intermediate states.

One further observes that the contributions of $1\gamma\times
2\gamma$ and $Z\times 2\gamma$ are negligible for $Q^2 \le$ 0.1
GeV$^2$. As $Q^2$ increases, the magnitudes of both contributions
increase  and become comparable in size with
$\delta_{\Delta}(\gamma\times\gamma Z)$ as $Q^2$ reaches 5.0
GeV$^2$. The cancelation between $1\gamma\times 2\gamma$ and
$Z\times 2\gamma$ contributions is also seen in Fig. {\ref{D}} with
the magnitude of $1\gamma\times 2\gamma$ contribution  larger than
that of $Z\times 2\gamma$ as in the $\delta_{N}$ case.

The $\gamma \times\gamma Z$ contribution exhibits more complicated
$Q^2$ and $\epsilon$ dependence. At lower $Q^2\le 0.1 GeV^2$, it
remains small until $\epsilon$ reaches  between $0.6 \sim   0.8$.
Then it increases rapidly before dropping drastically when
$\epsilon$ becomes very close to one. For $Q^2$ in the region of
$0.1\sim 1.0$ GeV$^2$, $\gamma ZE$ contribution is flat and almost
zero until $\epsilon$ increases past 0.8 and becomes  small and
negative  at forward angles. The behavior changes when $Q^2$ grows
larger than 1.0 GeV$^2$, as it  decreases monotonically with
increasing $\epsilon$, crosses zero at $\epsilon\sim 0.7$, and drops
rapidly as $\epsilon$ reaches 0.9.

To sum up, we see that at lower $Q^2\le$ 0.1 GeV$^2$, $\gamma ZE$
contribution  dominates.  When $Q^2$ reaches 5 GeV$^2$,
$1\gamma\times 2\gamma$ effect becomes dominant at backward angles
and brings the full $\delta_{\Delta}$ into negative. However, at
forward angles the $1\gamma\times 2\gamma$ contribution cancels the
sum of $1Z\times 2\gamma$ and $\gamma\times \gamma Z$ and the total
$\delta_{\Delta}$ becomes negligible.

\begin{figure}[h,b,t]
\centerline{\epsfxsize 4.0 truein\epsfbox{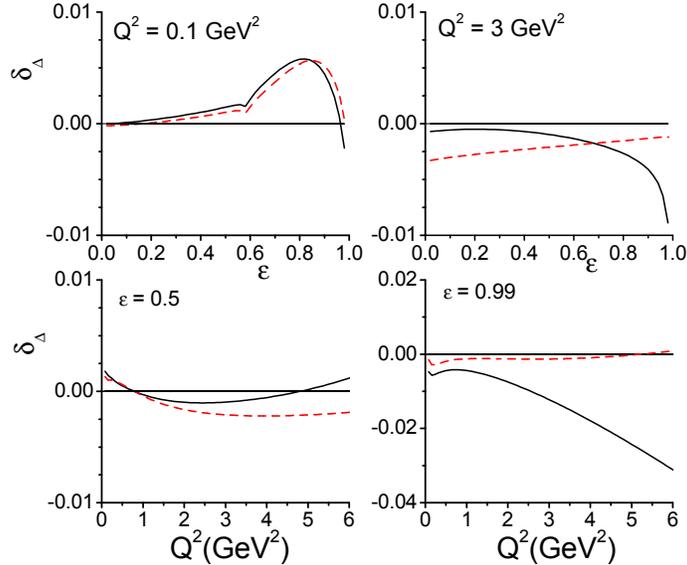}} \caption{The
short-dashed and solid lines  correspond to the results using vertex
function relations Eq. (\ref{theirs}) and  Eq. (\ref{ours}),
respectively.} \label{DV}
\end{figure}

Hereafter, we will restrict ourself to results of $\delta_\Delta$
obtained with form factors set B.

$\delta_{\Delta}$ has been calculated independently by
\cite{Nagata09} and \cite{Tjon09}. The  results are different
because of the following two reasons. The first is that different
relations between vertex functions of $\Gamma_{N\rightarrow \Delta}$
and $\Gamma_{\Delta\rightarrow N}$ are employed.  The other arises
from employing different values of $g_{3}$,  the Coulomb quadrupole
excitation strength of $N\rightarrow\Delta$.

In general, there are four diagrams associated with the $ \gamma ZE$
 diagram depicted in Fig. 1(d), two by interchanging the order of
 the exchanged $\gamma$ and $Z$ lines and two others
 from the associated cross-box diagrams. For simplicity, let's just
 consider only the two box diagrams without cross. We denote the
 amplitude of the diagram with $\gamma$ exchanged first by $M_{\gamma Z}$
 and the other with $Z$ exchanged first by $M_{Z\gamma}$, both with the
 $\Delta$ in the intermediate states and only the vector $ZN\Delta$ coupling.
 We may then write
\begin{eqnarray}
\delta_{\gamma Z}&=&\sum_{i,j=1}^3\tilde{g_j} g_i C_{ji}, \nonumber\\
\delta_{Z\gamma }&=&\sum_{i,j=1}^3 g_i \tilde{g_j} C_{ji}'.
\label{MM}
\end{eqnarray}
Our numerical results for the magnitudes of $C_{ji}$ agree
\cite{Tjon09a} with those obtained in \cite{Tjon09}. However, with
the use of the vertex relation of Eq. (\ref{theirs}), one would
obtain
\begin{equation}
C_{13}=-C'_{13},\,\,C_{23}=-C'_{23},\,\,C_{31}=-C'_{31},\,\,C_{32}=-C'_{32},\label{CCp}
\end{equation}
where the subscripts 1, 2, and 3 correspond to M1, E2, and C2
couplings, respectively, and $C_{ij}=C'_{ij}$ for the rest. On the
other hand, there would be no minus signs in Eq. (\ref{CCp}) if the
vertex relation of Eq. (\ref{ours}) is used. Consequently, after
summing up $\delta_{\gamma Z}$ and $\delta_{Z\gamma }$, the
crossing-couplings of C2 with M1 and E2 terms give no contribution
in the calculation of \cite{Tjon09}, while in \cite{Nagata09} no
cancelation between $\delta_{\gamma Z}$ and $\delta_{Z\gamma }$
occurs at all. Similar situation also takes place with the
amplitudes where the vertex $ZN\Delta$ is of axial-vector coupling.
The resulting discrepancy in the predictions for $\delta_{\Delta}$
are shown in Figs. {\ref{DV}} and {\ref{Dgc}}.

In the upper two figures of Fig. {\ref{DV}}, we show  the difference
arising from using vertex function relations
 of Eqs. (\ref{ours}) and (\ref{theirs}) for two fixed values of $Q^2=0.1$ and 3.0
 GeV$^2$. The solid and short-dashed lines correspond to the results using Eq.
 (\ref{ours}) and  Eq. (\ref{theirs}). At low
$Q^2$,  discrepancy is small. However at higher $Q^2$ the difference
between two results is significant at the forward angles. In the
lower two figures of Fig.  \ref{DV}, on the other hand, $\epsilon$
is fixed at 0.5 and 0.99, respectively, while $Q^2$ is varied. We
see large difference develop at large $Q^2$ in both cases.
Furthermore, when $\epsilon$ is fixed at 0.99, the solid line goes
downward but the dashed line goes upward, and when $Q^2$ reaches 6.0
GeV$^2$ the solid line goes down to about -0.04  but the dashed line
is almost zero.
\begin{figure}[h,b,t]
\centerline{\epsfxsize 4.0 truein\epsfbox{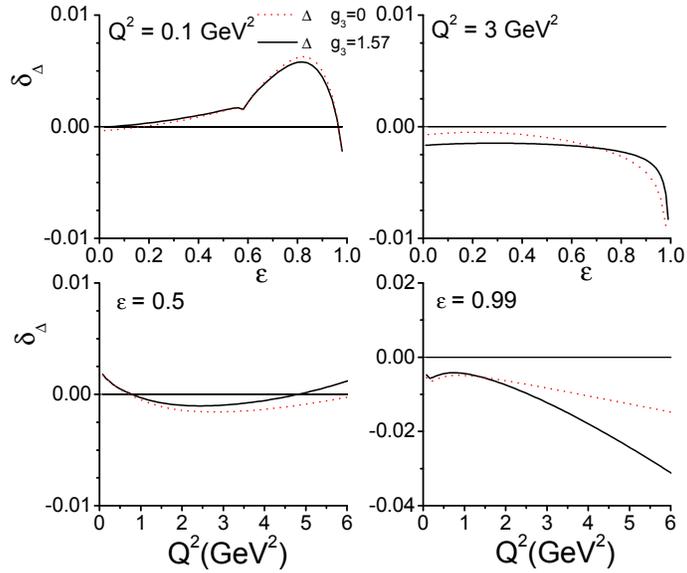}} \caption{The
dotted   and solid lines denote  results obtained with $g_{3}= 0$,
and  $g_{3}=1.57$, respectively.} \label{Dgc}
\end{figure}

In   Fig. {\ref{Dgc}}, the dotted and solid lines denote the results
obtained with $g_{3}=0$ and  $g_{3}=1.57$ as used in
\cite{Nagata09}, respectively. The difference at low $Q^2$ cases is
very small but becomes significant as $Q^2$ reaches 3.0 GeV$^2$,
especially at the forward angles. It underscores the important role
played by the Coulomb quadrupole transition in the evaluation of the
box diagrams with $\Delta$ intermediate states at high $Q^2$ and
large $\epsilon$.  In \cite{Tjon09},  $g_{3}$ (in our convention),
was varied from $-0.44$ to $1.28$ and  the effects of varying $g_3$
are found to be small. It is because the use of the vertex
relation of Eq. (\ref{theirs})  leads to cancelations  in the cross
couplings between  C2,  and M1 and E2 such that the effects of $g_3$
is reduced. Another reason is that only cases up to $Q^2\sim 1.0$
GeV$^2$ are explored.
\subsubsection{Total effects: sum of $\delta_{N}$ and $\delta_{\Delta}$}

Here we compare the behavior of $\delta_{N}$ and $\delta_{\Delta}$
and present their sum. We see from Fig. {\ref{ND}} that both of them
are sensitive to $Q^2$ and $\epsilon$. At $Q^2 =0.03$ GeV$^2$,
$\delta_{N}$ is dominant over $\delta_{\Delta}$ in the range $0\le
\epsilon \le 0.6$ because $\delta_{\Delta}$ is negligible there. As
$\epsilon$ further increases, $\delta_{\Delta}$ increases rapidly at
$\epsilon=0.6$ before dropping at extremely forward angles. These
behaviors are in sharp contrast with $\delta_N$ which simply
decreases as $\epsilon$ increases. The qualitative features of the
curves at $Q^2 =0.1$ GeV$^2$ remain the same with the one at $Q^2
=0.03$ GeV$^2$.

When $Q^2$  increases to 1.0 GeV$^2$, one sees that
$\delta_{\Delta}$   is very small and flat while  $\delta_{N}$
decreases monotonously with respect to $\epsilon$. As $Q^2$
increases up to 5 GeV$^2$, $\delta_{\Delta}$ becomes negative at
backward angles but becomes positive as $\epsilon$ increases. On the
contrary $\delta_{N}$ is always positive. We conclude that at small
$\epsilon$, $\delta_{N}$ is dominant but $\delta_{\Delta}$ becomes
important as $\epsilon$ grows.

Another way to compare $\delta_N$ with $\delta_\Delta$ is to see the
evolution of the $\delta's$ {\it w.r.t.} $Q^2$ at fixed $\epsilon$
as depicted in Fig. {{\ref{Q}}} for $\epsilon= 0.5$ and $\epsilon
=0.8$. The notation is the same as in  Fig. {\ref{ND}}. We clearly
see that for at $\epsilon =0.5$, $\delta_{\Delta}$ is small and of
opposite sign to $\delta_N$, while at larger value of $\epsilon =
0.8$, is always comparable with $\delta_N$ and becomes dominant at
large value of $Q^2$.

Since $\delta_N$ is substantially larger that $\delta_\Delta$ in the
region $\epsilon \le 0.8$, the total effect
$\delta=\delta_N+\delta_\Delta$  is very close to $\delta_N$. They
differ only after $\epsilon$ grows larger than $\sim 0.8$.

More quantitatively,  at $Q^2 =0.1$ GeV$^2$, the full correction,
i.e., the combined effect of $\delta_N$ and $\delta_\Delta$,
 reaches about 1.75\% at backward angle 135$^{\circ}$ (SAMPLE),
about 1.68\% at forward angle 35$^{\circ}$ (A4) and about -0.4\% at
very forward angle 6$^{\circ}$ (HAPPEX). On the other hand, when
$Q^2$ grows to 1.0 GeV$^2$, the  full correction starts from around
$1.4\%$ at backward angles and decreases to become less than
$-0.4\%$ at extreme forward angles.

\begin{figure}[h,b,t]
\centerline{\epsfxsize 4.0 truein\epsfbox{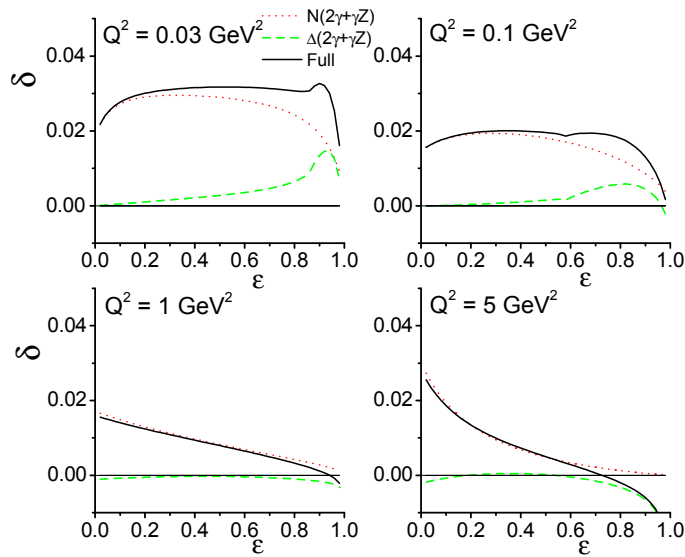}} \vskip-0.7cm
\caption{Two-boson exchange corrections with nucleon (dotted) and
$\Delta$ (dashed) intermediate states to $A_{PV}$, as functions of
$\epsilon$ from 0.1 to 0.9 at $Q^2 =0.03,\,\,0.1,\,\, 1.0,$ and
$5.0$ GeV$^2$, respectively. The solid lines denote their sums
$\delta=\delta_{N}+\delta_{\Delta}$. } \label{ND}
\end{figure}

\begin{figure}[h,b,t]
\centerline{\epsfxsize 4.0 truein\epsfbox{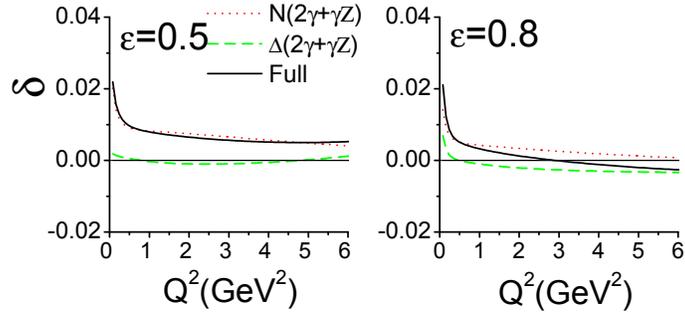}} \vskip-0.7cm
\caption{Two-boson exchange corrections with an intermediate nucleon
(dotted) and $\Delta$ (dashed) states to $A_{PV}$, as functions of
$Q^{2}$ from 0.1 to 6 GeV$^{2}$ at $\epsilon =0.5$ and 0.8,
respectively. The solid lines denote their sums
$\delta=\delta_{N}+\delta_{\Delta}$.} \label{Q}
\end{figure}

\subsubsection{Comparison with Marciano-Sirlin approximation}
Here we elucidate the relation between our results with those
obtained within  MS approximation \cite{Marciano84}.  Upon close
inspection, the method of MS actually contains three approximations.
The first one is to assume the momentum transfer
$Q=p_{1}-p_{3}=p_{4}-p_{2}=0$. Furthermore in the MS
approximation the electron mass is neglected and
$E_{lab}$ is taken to be zero. This is the second
approximation used by MS. Lastly, they take away the Coulomb
interactions because it was argued that its effect has been included
in the wave function of the bounded electron since they were
concerned with the atomic systems.

Moreover, the MS approximation includes no contribution from the
resonance intermediate states. Hence we shall compare our results
for $\delta_{N}(1\gamma\times\gamma Z)\equiv\bar\delta_N$ with values obtained in the MS approximation. Since
we have already seen  that the results do depend somewhat on the
form factors, we will employ the same form factors used in
\cite{Marciano84} in order to make the comparison more exact.


We first define some quantities to facilitate the comparison. In the
MS approximation, the parity asymmetry due to the $\gamma Z$
diagrams is given by,
\begin{equation}
A_{PV}^{MS,\gamma Z}(Q^2,E_{lab})=\frac{2Re[M^{(a)\dagger}M^{PV,MS}_{\gamma
Z}]}{|M^{(a)}|^{2}},
\label{APVMS}
\end{equation}
where $M^{(a)}$ is the $1\gamma E$  of Eq. (\ref{Born})
and  $M^{PV,MS}_{\gamma Z}$ the parity-violating part of the
$\gamma ZE$  amplitude evaluated within the MS approximation scheme.
The $Q^2$ and $E_{lab}$ dependence of Eq. (\ref{APVMS}) arises entirely from $M^{(a)}$ because
$M_{\gamma Z}^{PV,MS}$ is taken at $Q^2=0$ and $E_{lab}=0$.

\noindent We further introduce the following quantity,
\begin{equation}
\delta_{MS}(Q^2,E_{lab})=\frac{A_{PV}^{MS,\gamma Z}(Q^2,E_{lab})}{A_{PV}^{OBE}(Q^2, E_{Lab})}
=\frac{2Re[M^{(a)\dagger}M^{PV,MS}_{\gamma
Z}]/|M^{(a)}|^{2}}{2Re[M^{(a)\dagger}M^{(b)}]/|M^{(a)}|^{2}}=\frac{Re[M^{(a)\dagger}M^{PV,MS}_{\gamma
Z}]}{Re[M^{(a)\dagger}M^{(b)}]}.
\end{equation}

On the other hand,  the $\bar\delta_N$ we obtain is given as
\begin{equation}
\bar\delta_N(Q^2,E_{lab})=\frac{A_{PV}^{\gamma Z}(Q^2,E_{lab})}{A_{PV}^{OBE}(Q^2, E_{Lab})}
=\frac{2Re[M^{(a)\dagger}M^{PV,HM}_{\gamma
Z}]/|M^{(a)}|^{2}}{2Re[M^{(a)\dagger}M^{(b)}]/|M^{(a)}|^{2}}=\frac{Re[M^{(a)\dagger}M^{PV,HM}_{\gamma
Z}]}{Re[M^{(a)\dagger}M^{(b)}]},\label{deltaNbar}
\end{equation}
where $M^{(b)}$ is the $1Z E$ amplitude of Eq. (\ref{Born})
and $M^{PV,HM}_{\gamma Z}$  the parity-violating part of the $\gamma ZE$  amplitude
evaluated in our hadronic model with only the nucleon intermediate states included, with both
dependent on $Q^2$ and $E_{lab}$.
The relation between $\bar\delta_N$ and $\delta_{MS}$ is most transparent when  $Q^2=0,E_{lab}=0$ because
$M_{\gamma Z}^{PV,MS}$ is evaluated  at this point. In this limit,  $\delta_{MS}$ is given \cite{Marciano84} as,
\begin{equation}
\Delta_{MS}\equiv\delta_{MS}(Q^2=0,E_{lab}=0)=\rho_{\gamma Z}-\frac{4\kappa_{\gamma
Z}\sin^{2}\theta_W
}{1-4\sin^{2}\theta_{W}}=\frac{5\alpha_{em}}{2\pi}\left[K+\frac{4}{5}\xi_{B}\right],
\label{deltaMS}\end{equation}
where $\rho_{\gamma Z}$ and $\kappa_{\gamma Z}$ are
\begin{eqnarray}
\rho_{\gamma Z}&=&-\frac{2\alpha_{em}}{\pi}(1-4\sin^2\theta_W )\left[K+\frac{4}{5}\xi_{B}\right], \nonumber \\
\kappa_{\gamma Z}&=&-\frac{\alpha_{em}}{2\pi \sin^2\theta_W
}\left(\frac{9}{4}-4 \sin^2\theta_W \right)(1-4\sin^2\theta_W
)\left[K+\frac{4}{5}\xi_{B}\right]. \label{rho&kappa}
\end{eqnarray}
Here $K$ is the asymptotic contribution
obtained by carrying out the short-distance expansion in a
free-field theory. Its value is 8.58 if the onset scale is set to be
1 GeV. On the other hand, $\frac{4}{5}\xi_{B}$ corresponds to the
 the long-distance contribution of the $\gamma Z$ box diagram  estimated
in the Born approximation. Its value has been estimated to be 2.04.

Hence one obtains $\Delta_{MS}(low-k)=1.18 \%$ and
$\Delta_{MS}(high-k)=4.98\%$. It was argued  in \cite{Tjon09} that
the hadronic calculation as done here should correspond to the
so-called soft part because the form factors used in the hadronic
calculation function serves as a regulator and the contribution from
the higher loop momentum are suppressed. Accordingly, our result for
$\bar\delta_N(Q^2, E_{lab})$ should reproduce $\Delta_{MS}(low-k) =1.18 \%$ in the
proper MS limit as we discuss next.

In Fig. {\ref{CompMS}} we present our results for $\bar\delta_N(Q^2,E_{lab})$ by setting
$Q^2=0$ with varying $E_{lab}$. The full results and the Coulomb
contribution are denoted by the solid and short-dashed lines,
respectively. The difference between the solid and short-dashed
curves, represented by the long-dashed line, would correspond  to
the low-k contribution, to be compared with results of
\cite{Marciano84}. One sees that the long-dashed line, when $E_{lab}
$ goes to zero, does approach  1.18\%, a value given in  the MS
approximation if only low-k contribution is kept.
\begin{figure}
\centerline{\epsfxsize 4.0 truein\epsfbox{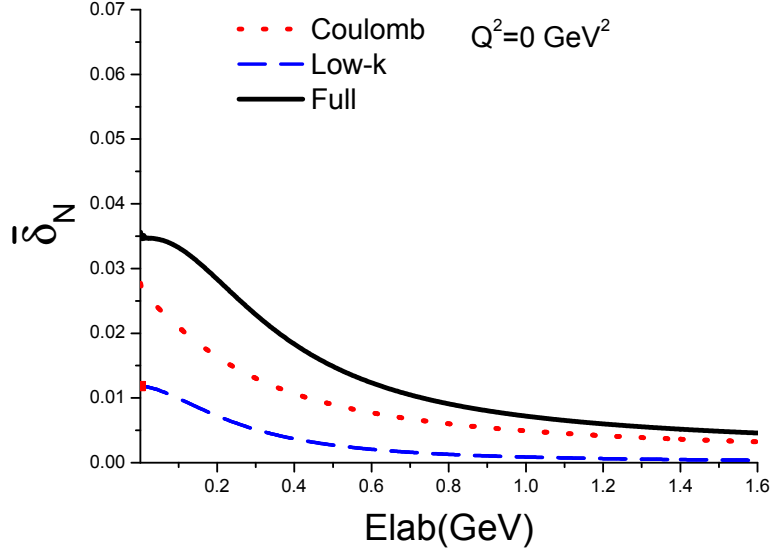}} \vskip-0.7cm
\caption{Comparison between our result for $\bar\delta_N\equiv\delta_N(1\gamma\times\gamma Z)$
of Eq. (\ref{deltaNbar}) at $Q^2=0$ with $\Delta_{MS}$ of Eq. (\ref{deltaMS})as obtained
within MS approximation. The dashed line denotes the contribution
corresponding to the low-k part of the MS approximation with  finite
$E_{lab}$. The Coulomb part is represented by the dotted line and
the total contribution is denoted as the solid line.} \label{CompMS}
\end{figure}

In other words, our calculation restores the value given by MS
approximation if we follow their scheme. On the other hand, it is
easy to see that the Coulomb interaction contribution is larger as
compared with the non-Coulomb part.
Furthermore the non-Coulomb contribution decreases more rapidly as
$E_{lab}$ increases. Note that the calculation in this section is
carried out at $Q^2=0$.
We see that the $\gamma ZE$ contributions is sensitive to $E_{lab}$
and it is necessary to go beyond the MS approximation.


\subsection{Extraction of the strange form factors}

Here we first examine the effects of the $2\gamma E$ and $\gamma ZE$
on the values of strange form factors extracted from HAPPEX
\cite{HAPPEX},   A4 \cite{A4}, and G0 experiments where data have
been taken at forward angles. In SAMPLE experiment, measurements of
both elastic $ep$ and  electron-deuteron ($eD$) scatterings are
combined to extract $G_M^s$. However, due to the fact that there is
no reliable way to estimate the TPE and $\gamma ZE$  contributions
to elastic $eD$ scattering, we do not know how to reanalyze SAMPLE
data. Naively, one may attempt to apply the simple hadronic model
here to the deuteron case. But as the deuteron is a loosely bound
system, treating   deuteron in a similar manner as proton is
questionable.

\subsubsection{The formulation of the extraction of the strange form factors}

All the existing analyses \cite{Young06, Liu07, Pate08}   extract
 strange from factors from Eq. (\ref{A123}) where the electroweak
radiative corrections are included in the parameters of $\rho$ and
$\kappa$ in the expression. The latest PDG values \cite{PDG2008} for
$\rho$ and  $\kappa$ are $\rho =0.9876$ and $\kappa =1.0026$. They
deviate from one because higher-order contributions like vertex
corrections, corrections to the propagators, and TBE effects are
taken into account. Since we have explicitly calculated the effect
of TBE effects in this study, we should   replace the contribution
of  TBE to the above-mentioned $\rho$ and $\kappa$ as estimated by
MS, with our results  to avoid double counting. Namely, one should
then subtract $\Delta \rho=\rho_{\gamma Z}$ and $\Delta
\kappa=\kappa_{\gamma Z}$ from $\rho$ and $\kappa$ and use
$\rho'=\rho-\Delta \rho$ and $\kappa'=\kappa-\Delta \kappa$ in Eq.
(\ref{A123}) instead.

As explained  earlier, $\rho_{\gamma Z}$ and $\kappa_{\gamma Z}$ in
Eq. (\ref{rho&kappa}) actually consist of two contributions, namely,
the high-k and low-k parts and our results  correspond  to the
  low-k part only. We should then only take away the soft
loop momentum contribution, which is associated with the $\xi_{B}$
term, and define $\Delta\rho,\Delta\kappa$ as follows:
\begin{eqnarray}
\Delta\rho&=&-\frac{2\alpha_{em}}{\pi}(1-4\sin^2\theta_W)\cdot\frac{4}{5}\xi_{B}=-0.73\times 10^{-3}\nonumber \\
\Delta\kappa&=&-\frac{\alpha_{em}}{2\pi
\sin^2\theta_W}\left(\frac{9}{4}-4
\sin^2\theta_W\right)(1-4\sin^2\theta_W)\cdot\frac{4}{5}\xi_{B}=-1.03\times10^{-3}.
\label{rho-kappa-soft}
\end{eqnarray}
Consequently, we  set the experimental parity asymmetry
$A_{PV}^{(Exp)}$ as follows:
\begin{eqnarray}
A_{PV}^{(Exp)}&\equiv &A_{PV}(1\gamma+Z+2\gamma+\gamma Z),
\nonumber\\&=&A_{PV}(\rho',\kappa')(1+\delta). \label{A-corrected}
\end{eqnarray}
With the value we obtain for $\delta$, we can determine
$A_{PV}(\rho',\kappa')$ and extract the strange form factors from
the resultant $A_2$ of Eq. (\ref{TreeA123}).

Furthermore, we introduce
\begin{equation}\overline{G}_E^s+\beta\overline{G}_M^s=(G_E^s+\beta
G_M^s)(1+\delta_G), \end{equation} to quantify the effects of the
$2\gamma E$ and $\gamma ZE$  to the extracted values of
$G^{s}_{E}+\beta G^{s}_{M}$, where $G_E^s+\beta G_M^s$ and
$\overline{G}_E^s+\beta\overline{G}_M^s$ are extracted from
$A_{PV}(\rho,\kappa)$ and $A_{PV}(\rho',\kappa')$, respectively.
From

\begin{eqnarray}
G^{s}_{E}+\beta G^{s}_{M}&=& \frac{\epsilon
(G^{\gamma,p}_{E})^2+\tau(G^{\gamma,p}_{M})^2}{a\rho\epsilon
G^{\gamma,p}_{E}}\left[A^{Exp}_{PV}-A_{1}(\rho,\kappa)-A_{3}\right],\nonumber
\\
\overline{G}^{s}_{E}+\beta \overline{G}^{s}_{M}&=& \frac{\epsilon
(G^{\gamma,p}_{E})^2+\tau(G^{\gamma,p}_{M})^2}{a\rho'\epsilon
G^{\gamma,p}_{E}}\left[\frac{A^{Exp}_{PV}}{1+\delta}-A_{1}(\rho',\kappa')-A_{3}\right],
\end{eqnarray}
we get,
\begin{eqnarray}
\delta_G&=&\frac{A^{Exp}_{PV}(\frac{\Delta\rho}{\rho}-\delta)+4a\rho\sin^{2}\theta_{W}
\Delta\kappa-\frac{\Delta\rho}{\rho} A_{3}}{A^{Exp}_{PV}-A_{0}}
\nonumber \\
&=&\left(\frac{-A^{Exp}_{PV}}{A^{Exp}_{PV}-A_{0}}\right)\delta
+\left(\frac{4a\rho\sin^{2}\theta_{W}}{A^{Exp}_{PV}-A_{0}}\right)
\Delta\kappa+\left(\frac{A^{Exp}_{PV}-A_{3}}{A^{Exp}_{PV}-A_{0}}\right)\frac{\Delta
\rho}{\rho}\nonumber \\
&=&\eta_{1}\delta+\eta_{2}\Delta \kappa+\eta_{3}\frac{\Delta
\rho}{\rho}, \label{deltaG}
\end{eqnarray}
where $A_{0}=A_{1}(\rho,\kappa)+A_{3}$. Note that the values of
$\eta_{1}$,   $\eta_{2}$, and $\eta_{3}$ all depend on the values of
the inputs of the nucleon form factors such as $G^{\gamma,
p}_{E,M}$, $G^{\gamma,n}_{E,M}$ and $G^{Z}_{A}$. As a result, the
value of $\delta_{G}$ also depends on those inputs.

We further define $\delta_{0}$, the corresponding
value of $\delta$ as would be obtained
in \cite{Marciano84} within $Q\equiv 0$ approximation scheme such
that $\delta_{G}=0$ if $\delta=\delta_{0}$. In other words,
difference between $\delta$ as we obtain and $\delta_0$, represents
the possible $Q^2$-dependence neglected in the estimation of
\cite{Marciano83}, such that $\delta_{G}$ vanishes when
$\delta=\delta_{0}$. Explicitly the value of $\delta_{0}$ is given
by
\begin{equation}
\delta_{0}=-\frac{\eta_{2}}{\eta_{1}}\Delta\kappa-\frac{\eta_{3}}{\eta_{1}}\frac{\Delta\rho}{\rho}.
\end{equation}
Obviously the value of $\delta_{0}$  depends on the inputs of the
proton and neutron electromagnetic form factors as well.

\subsubsection{Extraction of the strange form factors at HAPPEX, A4, and G0 experiments}

At  forward angles, $A_{3}$ in Eq. (\ref{A123}) is negligible
because both $\epsilon'=\sqrt{\tau(1-\tau)(1-\epsilon^2)}\ll 1$ and
$1-4\sin^{2}\theta_{W}\ll 1$.   It offers some advantages that the
strange form factors  can then be determined more accurately. This
is why many experiments, like HAPPEX, A4, and G0 are carried out at
very forward angles. In Table {\ref{tab1}}, we present our results
for $\delta_N,\, \delta_\Delta,\,$ their sum $ \delta$, besides
$\delta_0$ and $\delta_G$, for HAPPEX \cite{HAPPEX}, A4 \cite{A4},
and G0 \cite{G0} experiments. They are obtained with the use of form
factors set B. We also list the values of $G_{s}\equiv
G_{E}^{s}+\beta G_{M}^{s}$. For the G0 experiments, only
measurements of $A_{PV}$ are given in \cite{G0} and the
corresponding  values of $G_s$ listed in Table {\ref{tab1}} are
extracted by us with the use of the nucleon electromagnetic factors
parametrized in \cite{Alberico09}. The resultant change in the
values of $G_s$ after TBE effects are properly taken into account,
i.e., $\Delta G_{s}\equiv (\bar{G}_{E}^{s}+\beta
\bar{G}_{M}^{s})-(G_{E}^{s}+\beta G_{M}^{s})$, are given in the last
column of Table {\ref{tab1}}.

\begin{table}[htbp]
\begin{tabular}
{|c|c|c|c|c|c|c|c|c|c|} \hline Exp & $Q^2(GeV^2)$& $\epsilon$  &
$\delta_{N}(\%)$  & $\delta_{\Delta}(\%)$  & $\delta(\%)$ &
$\delta_{0}(\%)$ & $\delta_{G}(\%)$ & $G_{s}(10^{-2})$ & $\Delta G_{s} (10^{-2})$ \\
\hline HAPPEX & 0.477 & 0.974 &0.18&-0.27&-0.09&0.20&-2.54&1.4&-0.04\\
\hline HAPPEX & 0.109 & 0.994 &0.21&-0.80&-0.58&0.51&-20.63&0.7&-0.14\\
\hline G0& 0.122 & 0.9930 &0.21&-0.72&-0.51&0.61&-3.63&3.9&-0.14\\
\hline G0& 0.128 & 0.9926 &0.21&-0.70&-0.49&1.05&-1.28&9.2&-0.12\\
\hline G0& 0.136 & 0.9921 &0.21&-0.67&-0.44&0.81&-1.60&7.7&-0.12 \\
\hline G0& 0.144 & 0.9916 &0.20&-0.64&-0.41&0.38&14.14&-1.1&-0.16\\
\hline G0& 0.153 & 0.9911 &0.20&-0.61&-0.39&0.51&-3.50&3.8&-0.13 \\
\hline G0& 0.164 & 0.9904 &0.20&-0.58&-0.36&0.41&-9.18&1.5&-0.14\\
\hline G0& 0.177 & 0.9896 &0.20&-0.55&-0.32&0.31&6.19&-2.3&-0.14 \\
\hline G0& 0.192 & 0.9886 &0.19&-0.52&-0.29&0.35&-16.05&0.8&-0.12 \\
\hline G0& 0.210 & 0.9875 &0.19&-0.48&-0.29&0.30&48.25&-0.3&-0.14\\
\hline G0& 0.232 & 0.9860 &0.19&-0.44&-0.25&0.30&-20.25&0.6&-0.12\\
\hline G0& 0.262 & 0.9840 &0.19&-0.40&-0.21&0.35&-2.26&4.6&-0.10\\
\hline G0& 0.299 & 0.9814 &0.19&-0.36&-0.17&0.26&-8.68&1.2&-0.10 \\
\hline G0& 0.344 & 0.9783 &0.19&-0.32&-0.13&0.28&-1.99&4.4&-0.09 \\
\hline G0& 0.411 & 0.9735 &0.19&-0.27&-0.08&0.27&-1.18&6.4&-0.08 \\
\hline G0& 0.511 & 0.9657 &0.20&-0.23&-0.03&0.19&-2.10&2.8&-0.06\\
\hline G0& 0.628 & 0.9580 &0.21&-0.20&0.01&0.20&-0.71&6.8&-0.05 \\
\hline G0& 0.786 & 0.9413 &0.22&-0.18&0.04&0.15&-0.81&3.9&-0.03 \\
\hline G0& 0.997 & 0.9197 &0.25&-0.18&0.07&0.15&-0.32&7.6&-0.02\\
\hline A4& 0.108 & 0.83 &1.07&0.53&1.60&0.61&2.00&7.1&0.14\\
\hline A4& 0.23 & 0.83 &0.66&0.14&0.80&0.29&2.85&3.9&0.11 \\
\hline
\end{tabular}
\caption{The values of $\delta_{N},\delta_{\Delta}$, and their sum
$\delta$ for the HAPPEX \cite{HAPPEX}, G0 \cite{G0}, and A4
\cite{A4} data. We give the values of $\delta_{0}$, $\delta_{G}$,
$G_{s}$, and $\Delta G_{s}$ obtained with $g_3=1.57$, for those data
points.} \label{tab1}
\end{table}

All experimental data included in Table {\ref{tab1}} were obtained
in the near forward directions with $\epsilon\ge 0.8$. More
specifically, the HAPPEX and G0 data were taken at extremely forward
angles with $\epsilon\geq 0.92$. It is seen  from Table {\ref{tab1}}
that in this region there is a cancelation between $\delta_N$ and
$\delta_\Delta$ as they are of opposite sign. The magnitude of
$\delta_{\Delta}$ is always larger than $\delta_{N}$ at lower $Q^2
\le 0.7$ GeV$^2$. When $Q^2$ increases past 0.7 GeV$^2$,
$\delta_{N}$ overtakes $\delta_{\Delta}$ and the sum $\delta$
becomes positive. On the other hand, in the kinematical regions of
A4 data, both $\delta_{N}$ and $\delta_{\Delta}$  are positive such
that the sum $\delta$ is also positive.

Furthermore,  one notices that $\delta_{0}$ is in general larger
than $\delta_{N}$. It implies that MS  approximation  overestimates
the TBE contribution, besides neglecting the strong $Q^2$
dependence.

The values of $\delta_G$ presented in Table {\ref{tab1}} are
considerably smaller than what we reported  in \cite{Nagata09}. It
can be understood from the following reasons.  First, the values of
$\delta's$ listed are already different from before since they are
obtained with  different nucleon and $N\rightarrow\Delta$ form
factors. Previously in \cite{zhou07,Nagata09}, the nucleon form
factors used are of monopole type while the $N\Delta$ transition
form factors are taken to be of  dipole form. In addition,
$\rho_{\gamma Z}$ and $\kappa_{\gamma Z}$ obtained here are now used
to replace only the soft part contribution evaluated by MS, i.e.,
the $\Delta\rho$ and $\Delta\kappa$ of Eq. (\ref{rho-kappa-soft}),
as emphasized  in \cite{Tjon09}.   For example, for the HAPPEX data
at $Q^2=0.109$ GeV$^2$ and $\epsilon=0.994$, the value of
$\delta_{G}$ in \cite{Nagata09} is given as $-75.23\%$. However if we
use the value of $\delta=-0.58\%$ in Table {\ref{tab1}} instead of
the previous $\delta=-1.19\%$, then $\delta_{G}$ is reduced to
$-60.52\%$. If we further use the value  of $ \kappa_{\gamma
Z}=-1.03\times 10^{-3}$ identified as the hadronic contribution in
\cite{Marciano84} instead of the value of $-5.33\times 10^{-3}$,
then value of $\delta_{G}$ becomes $-14.99\%$. Lastly, using  the
value of $\Delta \rho=-0.73\times 10^{-3}$ given in Eq.
(\ref{rho-kappa-soft}), instead of the value of $-3.72\times
10^{-3}$, produces small change and leads to the final value of
$\delta_{G}=-20.63\%$ as given in Table I.

In general, the values of $\delta_{G}$ are smaller than 10\% and are
mostly negative with the exception of backward data of A4. For
HAPPEX data at $Q^2=0.109$ GeV$^2$ and G0 data at
$Q^2=0.144,\,0.192,\,0.210$ and 0.232 GeV$^2$, the magnitudes of
$\delta_G$ are large  and   range between $-20.95\%$ to $63.73\%$.
The magnitudes  of $\delta_{G}$ seem to behave irregularly. However
if one computes the $\Delta G_{s}\equiv (\bar{G}_{E}^{s}+\beta
\bar{G}_{M}^{s})-(G_{E}^{s}+\beta G_{M}^{s})$, the values of $\Delta
{G}_{s}$ are relatively stable with typical size of $-(0.1\sim
0.2)\times 10^{-2}$.   It is because those with large values of
$\delta_{G}$ have small  values of $G_{s}$.

Lastly, to illustrate the sensitivity of the corrections to the
extracted strange form factors, with respect to the possible
experimental uncertainties in the extracted value of $R_{SM}$ and
the resulting  Coulomb quardrupole excitation strength of the
$\Delta(1232)$, we give in Table II our results for $
\delta_\Delta,\, \delta\,\ \delta_0$, and $\delta_G$, obtained with
$g_3=0$ for some of the HAPPEX, A4, and G0 data. Comparison between
Tables I and II shows that the variations in the final corrections
to the extracted values of the strange form factors, when $g_3$
changes from 0 to 1.57, amount to about 20$\%$.

\begin{table}[htbp]
\begin{tabular}
{|c|c|c|c|c|c|c|} \hline Exp & $Q^2(GeV^2)$& $\epsilon$&
$\delta_{\Delta}(\%)$     &
 $\delta(\%)$
 & $\delta_{G}(\%)$  &$\Delta G_{s} (10^{-2})$  \\
\hline HAPPEX & 0.477 &0.974 &-0.30 &-0.13 &-2.81 &-0.04\\
\hline HAPPEX & 0.109  &0.994 &-0.94 &-0.73 &-23.36 &-0.16\\
\hline G0& 0.128 &0.9926 &-0.82 &-0.61 &-1.39 &-0.13\\
\hline G0& 0.144 &0.9916 &-0.75 &-0.55 &16.05 &-0.18\\
\hline G0& 0.164  &0.9904 &-0.68 &-0.48 &-10.32 &-0.15\\
\hline G0& 0.210  &0.9875 &-0.56 &-0.37 &54.38 &-0.16\\
\hline A4& 0.108  &0.83 &0.58 &1.65 &2.11 &0.15\\
\hline A4& 0.23  &0.83 &0.17 &0.83 &3.03 &0.12\\
\hline
\end{tabular}
\caption{The  values of $\delta_\Delta, \, \delta,\, \delta_G,$ and
$\Delta G_s$  obtained with $g_{3}=0$ for some of the HAPPEX, A4,
and G0 data.} \label{tab2}
\end{table}

\section{Summary}
In summary, we   present  the details of our calculation
\cite{zhou07,Nagata09} of the two-boson exchange effects in the
parity-violating $ep$ scattering within a simple hadronic model with
both the nucleon and $\Delta(1232)$ resonance intermediate states
included. We examine the sensitivity of the results with respect to
the form factors. We find that the nucleon contribution $\delta_N$
does show mild sensitivity to the form factors depending on whether
monopole or dipole form factors are used. However, little difference
is found between results obtained with a purely dipole form factors
set A and another more realistic form factors set B which differs
from set A only at higher $Q^2$. For the $\Delta$ contribution
$\delta_\Delta$, however, predictions obtained with the use of form
factors sets A and B do exhibit substantial difference at high
$Q^2$.

In addition, we compare our calculation \cite{Nagata09} for
$\delta_\Delta$ with a recent calculation of Ref. \cite{Tjon09}
where different relations relating vertex functions of
$\Gamma_{N\rightarrow \Delta}$ and $\Gamma_{\Delta\rightarrow N}$
are employed. Considerable discrepancy shows up at $Q^2 \geq 3.0$
GeV$^2$ and $\epsilon \geq 0.5$, when the Coulomb quardrupole
excitation (C2) strength of the $\Delta$, $g_3$ is nonvanishing.
Accordingly, if one takes $g_3=1.57$, a value determined from the
recent pion electroproduction data \cite{DMT} is used, our results
for $\delta_\Delta$ differ significantly with those given in
\cite{Tjon09}.

Furthermore, we clarify the relation between our results and the
well-known results of the $\gamma ZE$ effects given by Marciano and
Sirlin (MS).   We explicitly demonstrate that our calculation, with
only nucleon intermediate states included, restores the values given
by MS as long as we follow their scheme to set $Q\equiv 0$,
$E_{lab}=0$, and remove the Coulomb interaction.

We find that both the nucleon contribution $\delta_N$ and $\Delta$
contribution depend  on both $Q^2$ and $\epsilon$. $\delta_N$ is
always positive and decreases with increasing $\epsilon$. On the
contrary, $\Delta$ contribution $\delta_\Delta$ exhibits stronger
dependence on both $Q^2$ and $\epsilon$. In general, $\delta_N$
dominates over $\delta_\Delta$ except at extreme forward angles. The
sum $\delta=\delta_N+\delta_\Delta$ is then positive for $\epsilon
\le 0.95$ and turn negative after then.

We also present our result of the correction to the extracted values
of the strange form factors $G^{s}_{E}+\beta G^{s}_{M}$ from the
HAPPEX, A4, and G0 data at forward angles. Comparing with the
previous result \cite{zhou07, Nagata09}, the updated values are
reduced. However, the modification incurred in going beyond the MS
approximation  is still significant (up to $\sim 60\%$) for some
data. In addition, the sensitivity of the correction to the
extracted $G^{s}_{E}+\beta G^{s}_{M}$ values with respected to the
experimental uncertainty in the determination of $R_{SM}$ is found
to give rise to about $20\%$ variations when $R_{SM}$ changes from 0
to $-4.0\%$, or equivalently $g_3=0\sim 1.57$.

As we find significant contribution from TBE with $\Delta$
excitation in the extreme forward direction, where many of the
current experiments are performed,   question of the inclusion of
higher resonances comes up naturally. Naively, one would expect that
$\Delta(1232)$ would give the largest contribution since it is the
most prominent resonance at low energies. Higher resonances would be
suppressed because of their larger masses. However, only explicit
 calculation can answer this question.   Recent dispersion relation
calculation of the $\gamma ZE$ correction to $Q_W$
\cite{Gorchtein09} could be used to clarify this question in the
exact forward scattering. However, our results indicate that
$\delta$  depends sensitively with $Q^2$ at low momentum transfer so
whether dispersion relation method of \cite{Gorchtein09} can be
extended  to investigate the TBE correction to   strange form
factors remains to be further explored. Study of TBE effect with the
use of GPD as done in \cite{afanasev05} and \cite{Chen04} for TPE
effects, will also be very helpful in this regard.

\acknowledgements We acknowledge helpful communication with  J. A.
Tjon. This work is supported by the National Science Council of
Taiwan under grants nos. NSC096-2112-M033-003-MY3 (C.W.K.),
NSC098-2112-M002-006 (S.N.Y.) and by National Natural Science
Foundation of China under grant nos 10805009 (H.Q.Z). H.Q.Z. gladly
acknowledges the support of NCTS/HsinChu of Taiwan for his visit and
the warm hospitality extended to him by Chung Yuan Christian
University and National Taiwan University.

\end{document}